\documentclass{aa}  
\usepackage{graphicx}
\usepackage{txfonts}
\usepackage[hidelinks]{hyperref}
\usepackage{amsmath}	
\usepackage{amssymb}	
\usepackage[T1]{fontenc}
\usepackage{ae,aecompl}
\usepackage{multirow}
\usepackage{tabularx}
\usepackage{natbib}
\bibpunct{(}{)}{;}{a}{}{,} 

\usepackage{siunitx}

\begin{document}

 \title{ The broadband spectral analysis of 4U 1702-429\\ using XMM-Newton and BeppoSAX data}

   \subtitle{}

   \author{S. M. Mazzola\inst{1,2} \thanks{\email{simonamichela.mazzola@unipa.it}},  R. Iaria\inst{1},  T. Di Salvo\inst{1}, 
    M. Del Santo\inst{3},  A. Sanna\inst{2},  A. F. Gambino\inst{1},
   	 A. Riggio\inst{2},\\  A. Segreto\inst{3},  L. Burderi\inst{2}, A. Santangelo\inst{4}, N. D'Amico\inst{5}
   }

   \institute{Dipartimento di Fisica e Chimica,
 Universit\`a di Palermo, via Archirafi 36 - 90123 Palermo, Italy
         \and
         Dipartimento di Fisica, Universit\`a degli Studi di Cagliari, SP
Monserrato-Sestu, KM 0.7, Monserrato, 09042 Italy
   \and
  Istituto Nazionale di Astrofisica, IASF Palermo, Via U. La Malfa 153, I-90146 Palermo, Italy      
  \and
  IAAT Universität Tübingen, Sand 1, 72076 Tübingen, Germany
  \and
  INAF-Osservatorio Astronomico di Cagliari, via della Scienza 5, 09047 Selargius, Italy
  }

   \date{\today}

 
  \abstract
   {Most of the X-ray binary systems containing neutron stars classified as Atoll sources show two different spectral states, called soft and hard. Moreover, a large number of these systems show a reflection component relativistically smeared in their spectra, which gives information on the innermost region of the system.} 
  {Our aim is to investigate the poorly studied broadband spectrum of the low mass X-ray binary system 4U 1702-429, which was recently analysed combining {\it XMM-Newton} and {\it INTEGRAL} data. The peculiar value of the reflection fraction brought us to analyse further broadband spectra of 4U 1702-429.}
  {We re-analysed the spectrum of the {\it XMM-Newton/INTEGRAL} observation of 4U 1702-429 in the 0.3-60 keV energy range and we extracted three 0.1-100 keV spectra of the source analysing three observations collected with the \textit{BeppoSAX} satellite.}
  {We find that the {\it XMM-Newton/INTEGRAL} spectrum is well fitted using a model composed of a disc blackbody plus a Comptonised component and a smeared reflection component. We used the same spectral model for the {\it BeppoSAX} spectra, finding out that the addition of a smeared reflection component is statistically significant. The best-fit values of the parameters are compatible to each other for the {\it BeppoSAX} spectra.
We find that the reflection fraction is $0.05^{+0.03}_{-0.01}$ for the {\it XMM-Newton/INTEGRAL} spectrum and between 0.15 and 0.4 for \textit{BeppoSAX} ones.}
  {The relative reflection fraction and the ionisation parameter are incompatible between the {\it XMM-Newton/INTEGRAL} and the {\it BeppoSAX} observations and the characteristics of the Comptonising corona suggest that the source was in a soft state in the former observation and in a hard state in the latter.}

\authorrunning{S. M.\ Mazzola et al.}

\titlerunning{The broadband spectral analysis of 4U 1702-429}

   \maketitle
%

\section{Introduction}
Low Mass X-ray Binaries hosting neutron stars (hereafter NS-LMXBs) are
binary systems in which a weakly magnetised neutron star (NS) accretes matter from a
low mass ($< 1$ M$_{\odot}$) companion star via the Roche-lobe overflow.
In the X-ray spectra of these sources we can distinguish some main components: a soft thermal component due to the blackbody emission from the NS and/or the accretion disc, a hard component due to the Comptonisation of soft photons from a hot electron corona located in the innermost region of the system, and the reflection component originated by the interaction of photons outgoing the Comptonising cloud scattered by the surface of the disc.
The emission of a NS-LMXB is characterised by two spectral states, called soft and hard. In the soft state (SS) the continuum spectrum can be described by a (predominant) blackbody or disc-blackbody component and a harder saturated Comptonised component \citep[see e.g.][]{DiSalvo_09,DiSalvo_05,Piraino_07,Barret_02}. Whilst the hard state (HS) spectrum can be described by a weak thermal emission, sometimes not significantly detected \citep[see e.g.][]{Ludlam_16}, plus a power law with a high energy cut-off explained as due to inverse Compton scattering of soft photons in the hot electron corona \citep[see e.g.][]{DiSalvo_15,Dai_10,Cackett_10}. Some HS spectra have been described in terms of a hybrid model of a broken power law/thermal Comptonisation component plus two blackbody components \citep{Lin_07,Armas_17}. The use of a broken power law model, however, does not allow to obtain information about the origin of the seed photons of the Comptonised component, i.e. it does not allow to distinguish between the contribution to the Comptonised component of a blackbody or a disc-blackbody. In this sense, a Comptonisation model is more useful in order to identify the origin of all the spectral components, and in particular the main responsible for the soft thermal emission. Recently, some HS spectra have been interpreted in terms of a double Comptonised component with seed photons emitted by NS and the accretion disc \citep{Zhang_16}.

The broad-band spectral analysis is useful in order to get information on the nature of the compact object (a BH or a NS) present in a LMXB, as well as to infer information on the innermost region of the system \citep[e.g.][]{DiSalvo_06}. Of particular interest, in this sense, is the study of the reflection component, that originates from direct Compton scattering of the Comptonized photons outgoing the hot corona with the cold electrons in the top layers of the inner accretion disc. In most of the cases it results in the so-called Compton hump around 20-40 keV \citep{Egron_13,Miller_13,Barret_00}. The reflection component can show also some discrete features due to the fluorescence emission and photoelectric absorption by heavy ions in the accretion disc. The reflection is more efficient in SS, usually showing a higher degree of ionisation of the accreting matter ($\xi >500$) and the presence of stronger features with respect to HS. 
This is an evidence both in NS-LMXBs \citep{DiSalvo_15,Egron_13} and in  BH-LMXBs \citep{Done_07}. The efficiency of the reflection is mainly indicated by the presence of a strong emission line from highly ionised Fe atoms; this probably reflects a greater emissivity of the accretion disc, which is proportional to $r^{-betor}$, where $r$ is the radius of the disc at which the incident radiation arrives and $betor$ is the power-law dependence of emissivity and it is usually found to be between 2 (in the case of a dominating central illuminating flux) and 3 (which describes approximately the intrinsic emissivity of a disc). The higher emissivity in SS is a consequence of a disc (generally) closer to the NS surface than in HS.
Indeed broad emission lines (FWHM up to 1 keV) in the Fe-K region (6.4 - 6.97 keV) are often observed in the spectra of NS-LMXBs with both an inclination angle lower than 60$^{\circ}$ \citep[e.g.][]{Matranga_17,Chiang_16,Papitto_13,Sanna_13,Piraino_12,Cackett_09, DiSalvo_09,Iaria_09, Shaposhnikov_09,Pandel_08,Cackett_08} and with an inclination angle between $\ang{60}$ and $\ang{90}$
\citep[the so-called dipping and eclipsing sources, e.g.][]{Ponti_18,Pintore_15,Diaz_12,Diaz_09,Iaria_07,Boirin_05,Parmar_02,Sidoli_01}. 
These lines are identified with the K$\alpha$ radiative transitions of iron at different ionisation states.
These broad lines most likely originate in the region of the accretion disc close to the compact object where matter is rapidly rotating and reaches velocities up to a few tenths of the speed of light. Hence the whole reflection spectrum is believed to be modified by Doppler shift, Doppler broadening, and gravitational redshift \citep[see e.g.][]{Fabian_89,Laor_91}, which produce the characteristic broad and skewed line profile. 

It should be mentioned that relativistic (plus Compton) broadening is not the unique physical interpretation for the large width of the iron line profile in these systems. For instance, \cite{Tit_03} suggested that the characteristic asymmetric skewed profile of the iron line could originate from an optically thick flow ejected from the disc, outflowing at relativistic speeds. However, for an increase of the observed broadening this model requires a corresponding increasing of the equivalent hydrogen column, due to the increased number of scatterings, and this correlation seems to be not present for NS-LMXBs \citep{Cackett_13}.
The study of the line profile gives therefore important information on the ionisation state, geometry and velocity field of the reprocessing plasma in the inner accretion flow.

In this paper we report on a broad-band spectral analysis of 
the X-ray source 4U 1702-429 (Ara X-1) that is a NS-LMXB showing type-I X-ray bursts. The source was detected as a burster with {\it OSO 8} \citep{Swank_76}. \cite{Oost_91} classified 4U 1702-429 as an atoll source using {\it EXOSAT} data. \cite{Galloway_08}, analysing the photospheric radius expansion of the observed type-I X-ray bursts,
inferred a distance to the source of $4.19 \pm 0.15$ kpc and $5.46 \pm 0.19$ kpc for a pure hydrogen and pure helium companion star, respectively. \cite{Mark_99}, using the data of the proportional counter array (PCA) onboard the {\it Rossi X-ray Timing Explorer (RXTE)} satellite, detected burst oscillations at 330 Hz that could be associated with the spin frequency of the NS.

Using {\it Einstein} data, \cite{Christian_97} modelled the continuum emission adopting a cut- off power law with a photon-index between 1.3 and 1.5 and a cut-off temperature between 8 and 16 keV.  \cite{Mark_99}, analysing {\it RXTE/PCA} data, adopted the same model and obtained a temperature between 3.5 and 4.6 keV. Using {\it XMM-Newton} and {\it INTEGRAL} spectra, \cite{Iaria_16} revealed the presence of a broad emission line at 6.7 keV. These authors fitted the spectra adopting a disc blackbody component plus Comptonisation and reflection component from the accretion disc. An inner disc temperature of 0.34 keV, an electron temperature of 2.3 keV and a photon index of 1.7 - both associated with the Comptonising cloud - have been obtained. The equivalent hydrogen column density associated with the interstellar medium was $2.5 \times 10^{22}$ cm$^{-2}$ and the ionisation parameter associated with the reflecting plasma was $\log (\xi)=2.7$.

In order to compare recent observations with previous ones and explore different spectral states of the source (in particular, concerning the presence of the reflection component), we re-analysed the {\it XMM-Newton/INTEGRAL} spectrum of 4U 1702-429 already studied by \cite{Iaria_16}. We also present for the first time the analysis of three 0.1-100 keV {\it BeppoSAX} spectra, showing that the addition of a reflection component is statistically required.

\section{Observations and data reduction}
The narrow field instruments (NFIs) on board the \textit{BeppoSAX}
satellite observed 4U 1702-429 three times between 1999 and 2000. The NFIs are four co-aligned instruments which cover three decades in energy, from 0.1 keV up to 200 keV. The Low-Energy Concentrator Spectrometer \citep[LECS, operating in the range 0.1-10 keV;][]{Parmar_97} and the Medium-Energy Concentrator Spectrometer \citep[MECS, 1.3-10 keV;][]{boella_97} have imaging capabilities with fields of view (FOV) of 20$\arcmin$ and 30$\arcmin$ radii, respectively. We selected data in circular regions - centred on the source - of 8$\arcmin$ and 6$\arcmin$ radii for LECS and MECS, respectively. The background events were extracted from circular regions with the same radii adopted for the source-event extractions and centred in a detector region far from the source. The High-Pressure Gas Scintillator Proportional Counter \citep[HPGSPC, 7-60 keV;][]{manzo_97} and Phoswich Detector System \citep[PDS, 13-200 keV;][]{frontera_97} are non-imaging instruments, with a FOVs of $\sim 1^{\circ}$ FWHM delimited by collimators. The background subtraction for these instruments is obtained using the off-source data accumulated during the rocking of the collimators.
  \begin{figure*}
   \centering
   \includegraphics[scale=0.45]{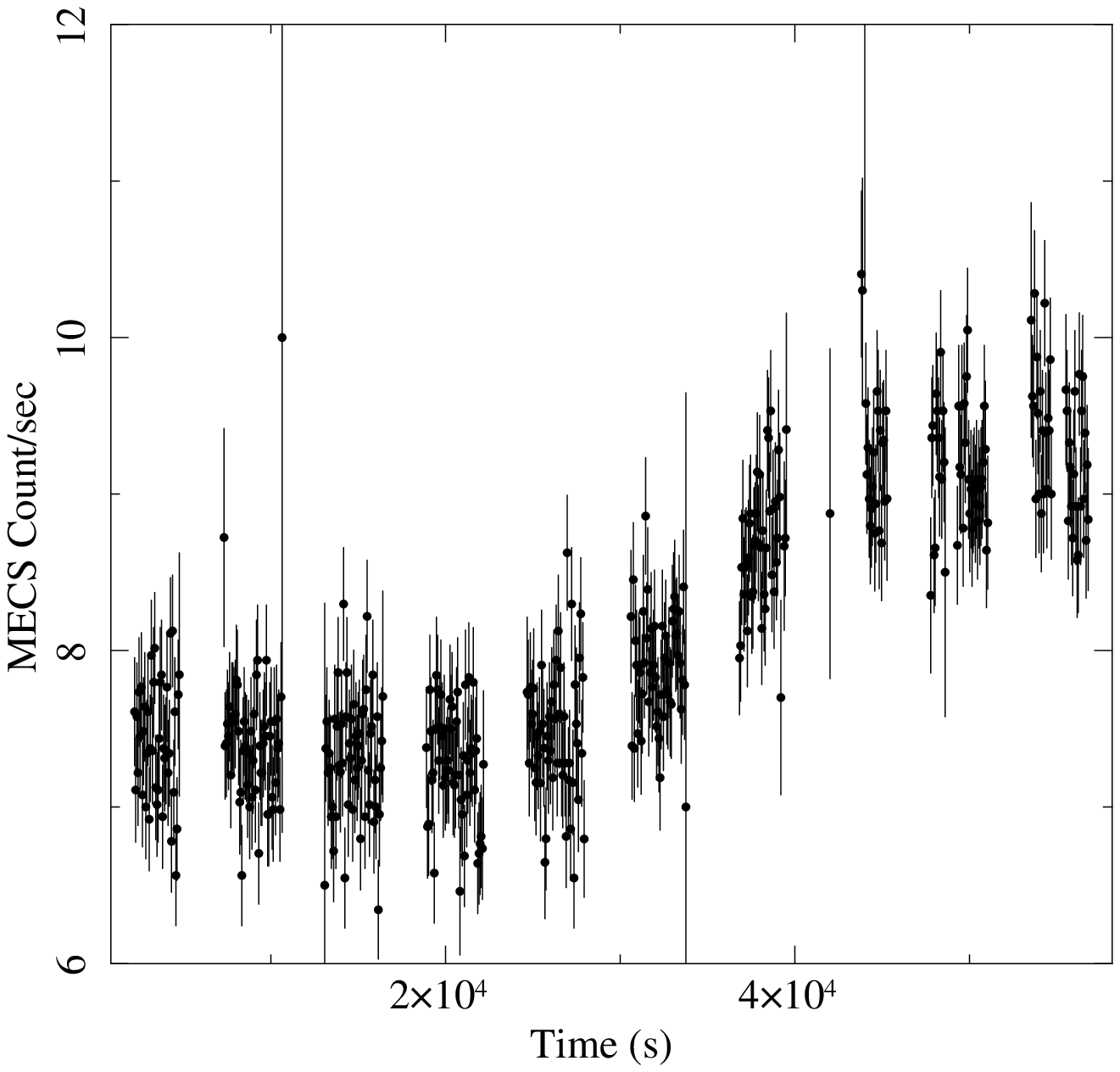}
   \includegraphics[scale=0.45]{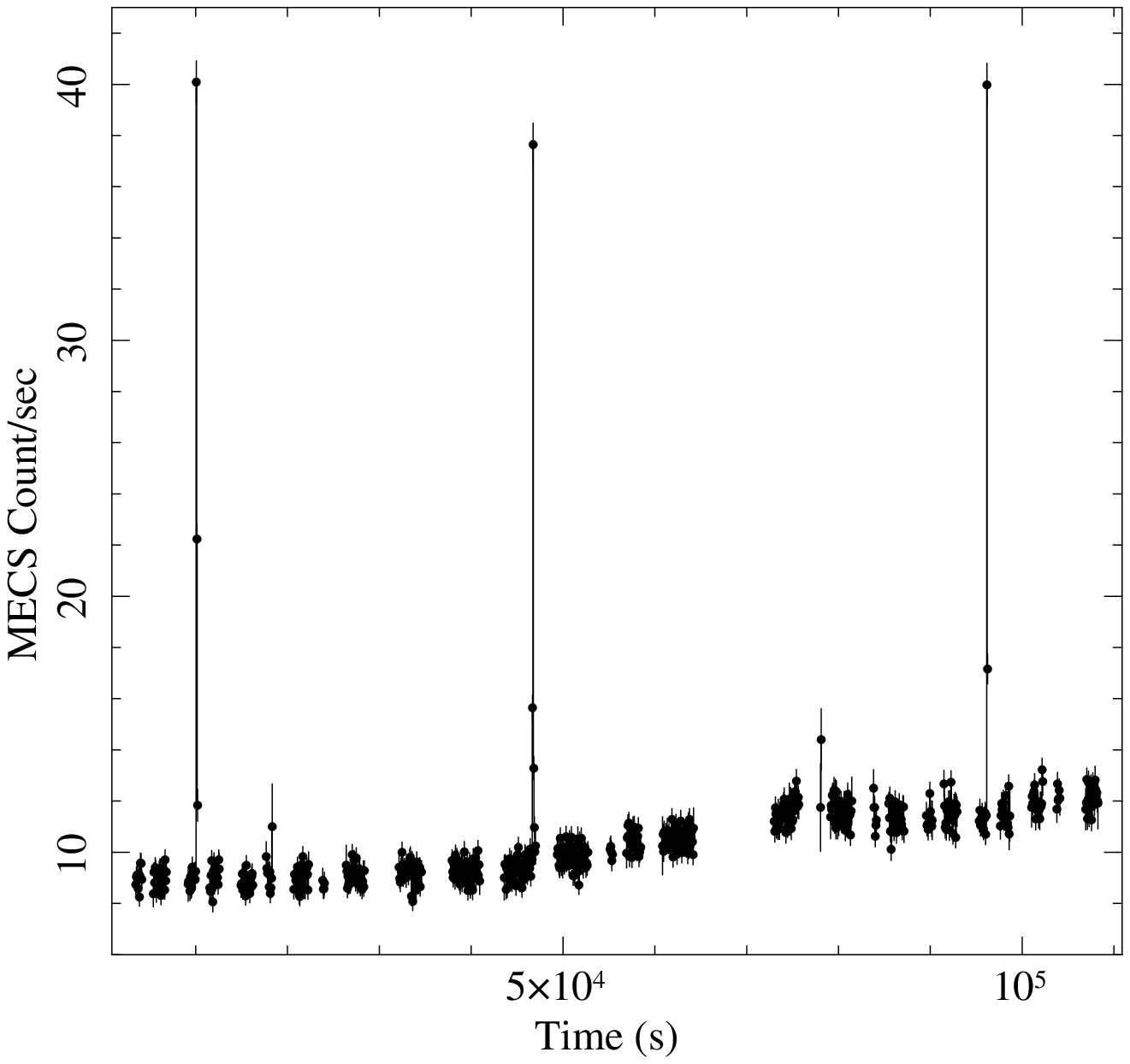}
   \includegraphics[scale=0.45]{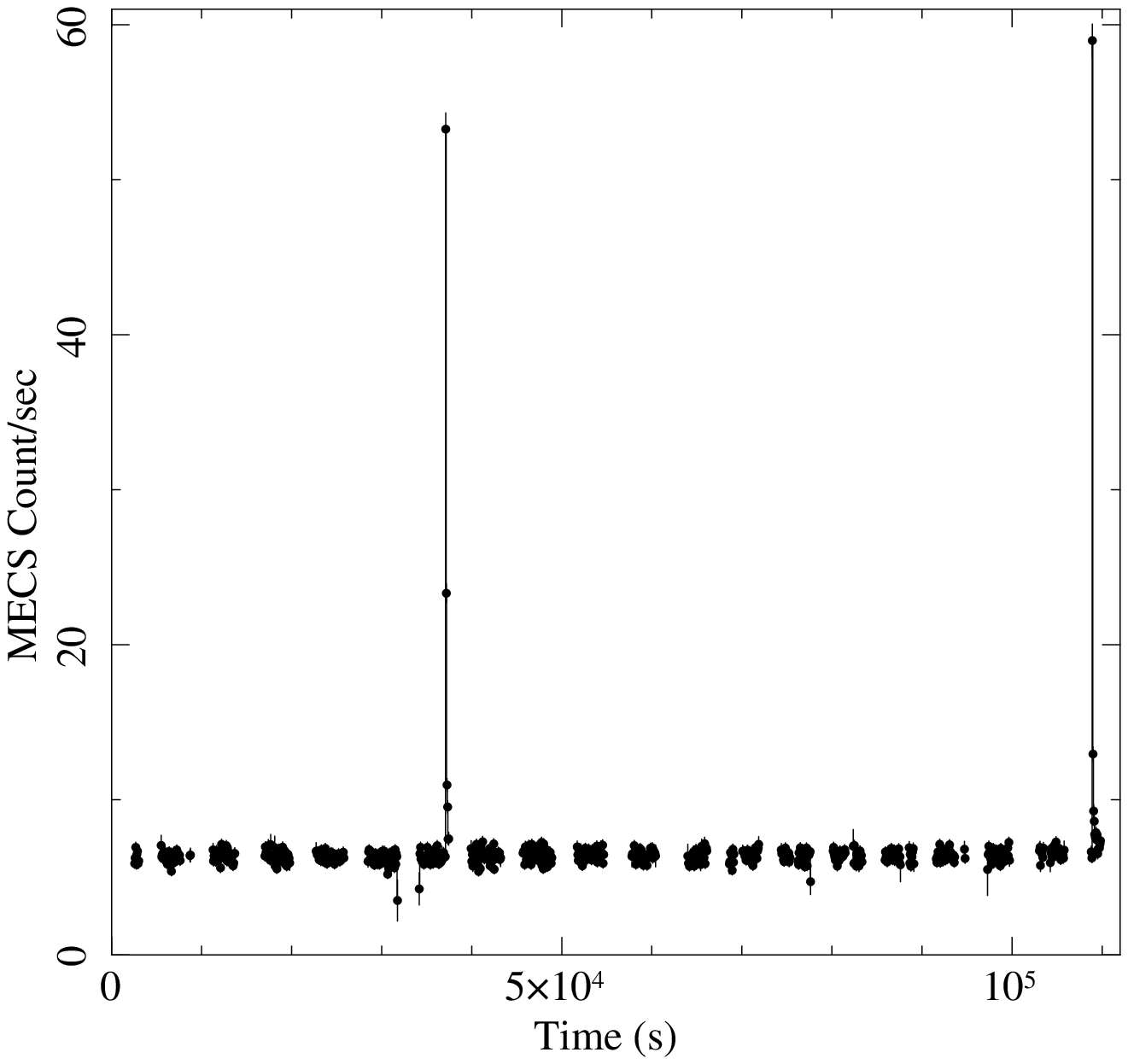}
   \caption{MECS23 light curves of the source 4U 1702-429 for the
  observations A, B and C in left, central and right panel,
  respectively. The bin time is 64 s. Three and two X-ray Type-I bursts occurred during observations B and C, respectively.}
              \label{fig:lc_MECS}
    \end{figure*}
  \begin{figure*}
   \centering
   \includegraphics[scale=0.45]{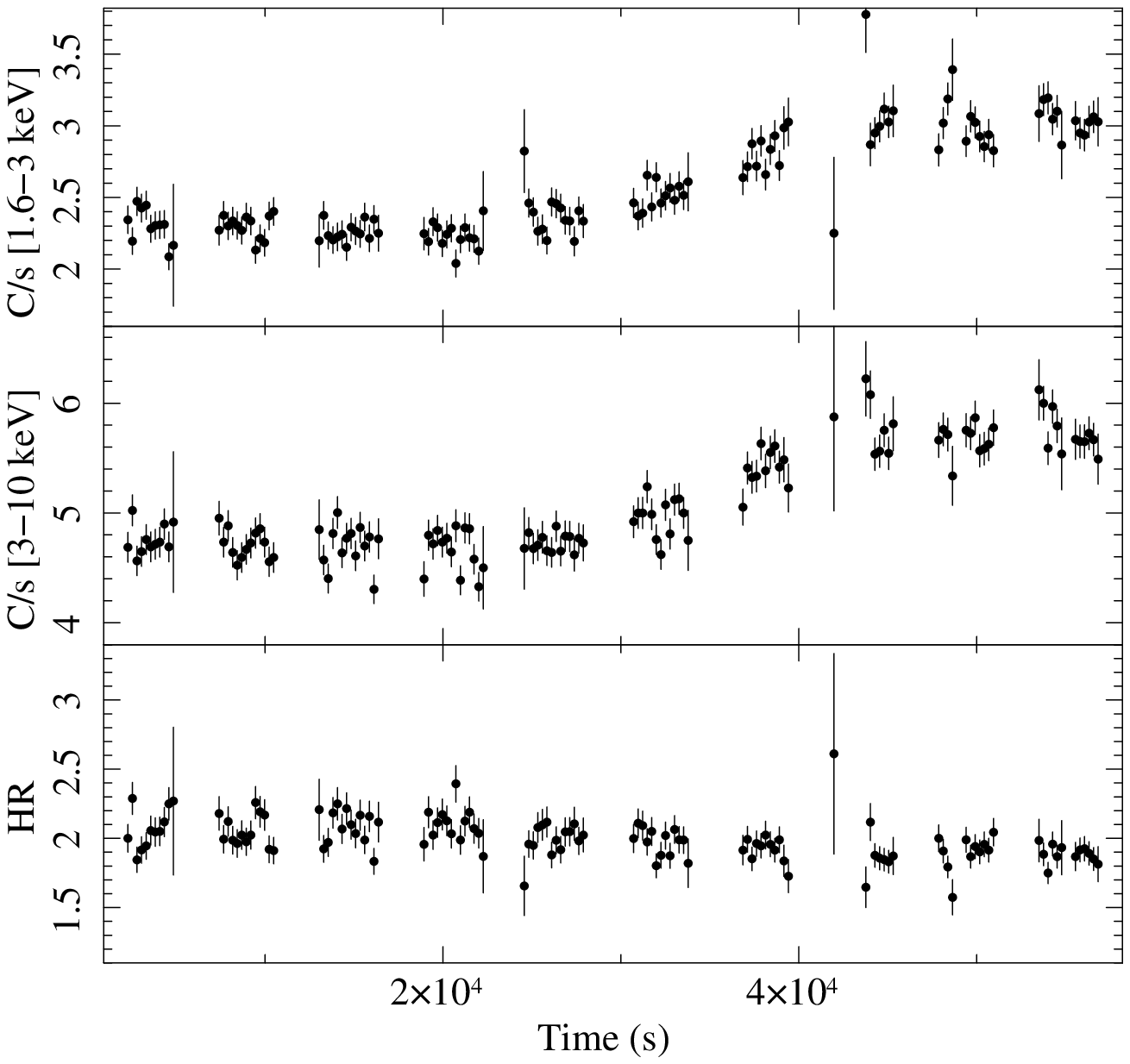}
   \includegraphics[scale=0.45]{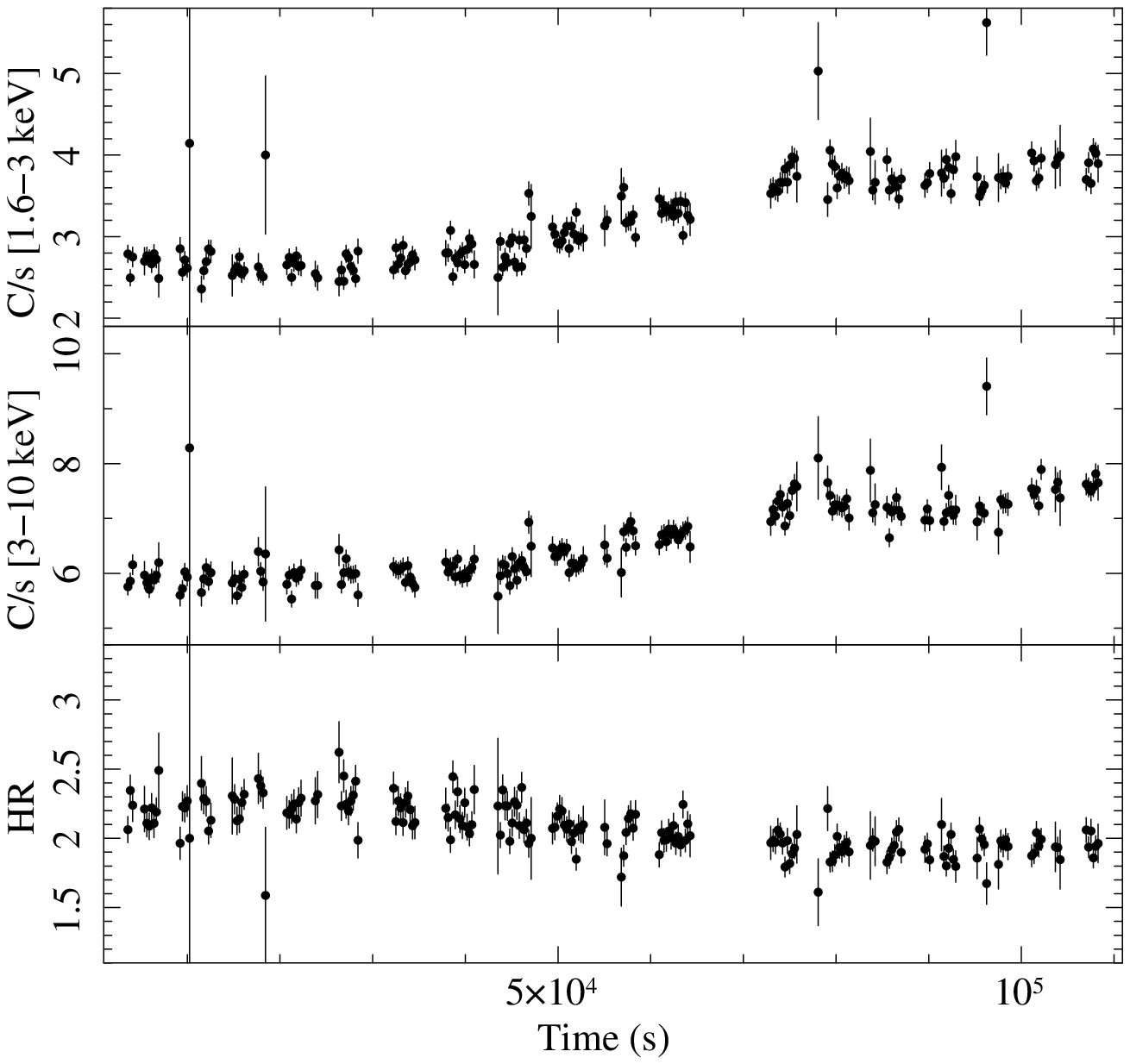}
   \includegraphics[scale=0.45]{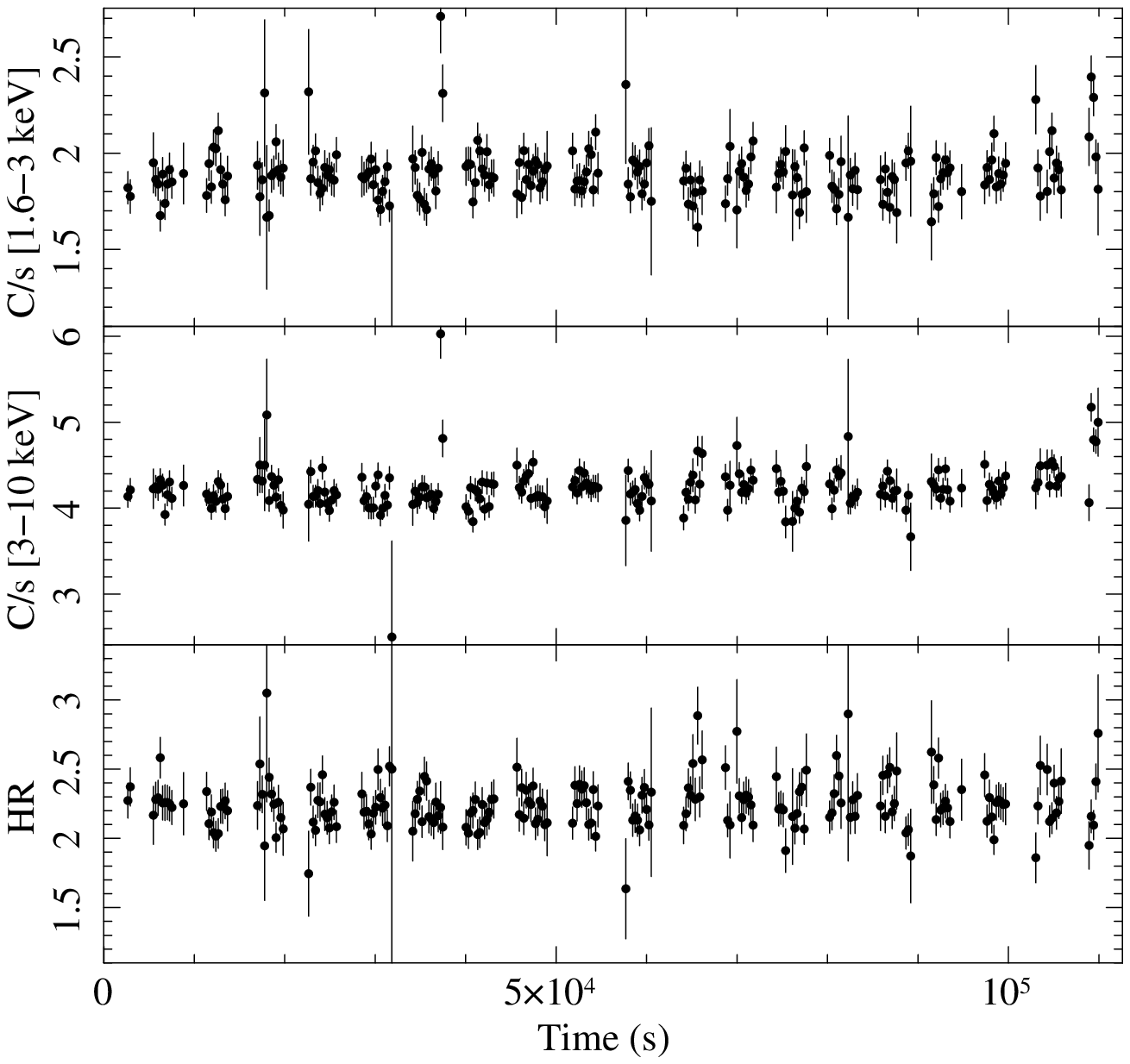}
   \caption{Left panel: MECS23 light curves and HR for obs. A.
   Central panel: MECS23 light curves and HR for obs. B. Right
  panel: MECS23 light curves and HR for obs. C. In each panel,
  from the top to the bottom: MECS23 light curve in the energy
  band 1.6-3 keV, in the energy band 3-10 keV and the
  corresponding Hardness Ratio. The bin time is 256 s.}
              \label{fig:HR_MECS}
    \end{figure*}

MECS is composed of three modules but only MECS2 and MECS3 were active during these observations. The event files of these two instruments are merged and we indicate them as MECS23. The first observation (obsid. 2069400100, hereafter observation A) was performed between 1999 February 27 17:34:34 UT and February 28 09:12:06 UT for a duration time of 58 ks, the second observation (obsid. 2122400100, hereafter observation B) was performed between 2000 August 24 19:22:51 UT and August 26 01:04:01 UT for a duration time of 104 ks, and finally the third observation (obsid. 2122400200, hereafter observation C) was taken between 2000 September 23 22:41:50 UT and and September 25 04:30:29 UT for a duration of 107 ks.
%
\begin{table} [!htbp]
\caption{The NFI's exposure times for each of the three observations of 4U 1702-429}
\label{Tab_exp}
\centering
\scriptsize
\begin{tabular}{c c c c}
\hline
\hline
   & \multicolumn{3}{c}{Exposure time (ks)}\\
Instrument & Observation A & Observation B &Observation C \\
\hline
LECS & 7.1 & 19.1 & 16.3 \\
MECS23 & 27.5 & 40.4 & 46.8\\
HPGSPC & 28.3 & 40.1 & 41.8 \\
 PDS & 13.5 & 20.1 & 20.9 \\

\hline
\hline
\end{tabular}
\end{table}
\begin{table} [!htbp]
\caption{Time intervals excluded for bursts removing}
\label{tab:burst}
\centering
\scriptsize
\begin{tabular}{c c c}
\hline 
\hline
&  \multicolumn{2}{c}{Excluded temporal interval (in ks from the start time)}\\
Instrument & Observation B & Observation C\\
\hline
LECS &  41.160 - 41.350 & \\
\\
\multirow{ 3}{*}{MECS23} 
&  6.630 - 6.750 &  34.570 - 34.750 \\
 &  43.250 -43.350 & 106.350 - 106.550 \\
& 92.680 - 92.760 & \\
\\
\multirow{ 3}{*}{HPGSPC} 
 & 6.800 - 7.000 & 34.430 - 34.560 \\
& 43.400 - 43.600 & 106.240 - 106.500 \\
&   92.800 - 93.100 & \\
\\
\multirow{ 2}{*}{PDS} 
& 45.340 - 45.354 & 34.530 - 34.554 \\
& 94.767 - 94.780 & 106.316 - 106.334 \\

\hline
\hline
\end{tabular}
\end{table}
We used several tools of SAXDAS 2.3.3 and HEASOFT 6.20 to extract scientific products from clean event files, downloaded from the Multi-Mission Interactive Archive at the ASI Space Science Data Center (SSDC). Initially, we extracted MECS23 light curve in the 1.3-10 keV energy range for each observation. We show the three MECS light curves with a bin time of 64 s in \autoref{fig:lc_MECS}. During obs. A the count rate is constant at 8.5 c/s from the start time up to 35 ks, when the count rate increases up to 10 c/s (\autoref{fig:lc_MECS}, left panel). During obs. B three type-I X-ray bursts are present at 6.6 ks, 43.3 ks and 92.6 ks from the start time. The count rate at the peak of the bursts is 40 c/s, 35 c/s and 40 c/s, respectively, whilst the count rate of the persistent emission gradually increases from 9.5 c/s to 11 c/s (\autoref{fig:lc_MECS}, middle panel). During obs. C two type-I X-ray bursts are present at 34 ks and 106 ks from the start time. The count rate at the peak of the bursts is 52 c/s and 60 c/s, respectively. The count rate of the persistent emission is constant at 8 c/s (\autoref{fig:lc_MECS}, right panel). \\
LECS exposure times are 7.1 ks, 19.4 ks and 12.3 ks for obs. A, B, and C, respectively. LECS light curves corresponding to obs. A and C do not show type-I X-ray bursts, while the LECS light curve corresponding to obs. B shows only the first type-I X-ray burst
observed in the MECS23 light curve of the same observation.
The exposure times of each instrument for the three observations are shown in \autoref{Tab_exp}.

Since our aim is the study of the persistent emission of the source, we excluded the bursts detected from the light curves. Using the {\tt XSELECT} tool, we determined the time intervals at which bursts occurred and excluded them (\autoref{tab:burst}). The same operation was carried out on the background light curves to keep the same exposure time.
 
Once bursts time intervals were excluded, we build the MECS23 hardness ratio (HR): we divided MECS data in the energy bands 1.6-3 keV and 3-10 keV and produced HR for each observation. We show the 1.6-3 keV light curves, the 3-10 keV light curves and the corresponding HRs in \autoref{fig:HR_MECS} for obs. A, B and C. The HRs are quite constant along the whole observations, then we extracted a spectrum of the persistent emission for each of the three ones.

We extracted MECS and LECS spectra using {\tt XSELECT}, and we grouped each ones in order to have at least 25 photons for each energy channel.
We used the {\tt hpproducts} and {\tt pdproducts} tools to obtain the HP and PDS background-subtracted spectra, cleaned from the bursts. The PDS spectra were grouped adopting a logarithmic rebinning. 

The combined {\it XMM-Newton} and {\it INTEGRAL} persistent spectrum of 4U 1702-429 used in this work was obtained as described by \cite{Iaria_16}. The exposure times are 37 ks for the reflection grating spectrometers
\citep[RGS,][]{Herder_01}, 36 ks for the spectra collected by the two
MOS charge-coupled devices (CCDs) \citep[MOS1 and MOS2,][]{Turner_01}
and by the pn CCDs \citep[PN,][]{Struder_01} of the European Photon
Imaging Camera (EPIC). The MOS1 and MOS2 spectra were combined
togheter as well as the RGS1 and RGS2 spectra. The exposure time of
JEM-X2 \citep{Lund_03} and IBIS/ISGRI \citep{Ubertini_03,Lebrun_03} are 6 ks and 130 ks, respectively.

\section{Spectral analysis}
We used the {\tt XSPEC} software package v12.9.1 to fit the spectra.
For all observations we fitted the continuum direct emission with a model composed of a multicolour disc blackbody emission \citep[{\tt diskbb} in {\tt XSPEC}, ][]{Mitsuda_84,Makishima_86} plus a thermal Comptonisation \citep[{\tt nthComp} in {\tt XSPEC},][]{Zdi_96,Zichi_99}, in which the {\tt inp\_type} parameter was set to 1, indicating that the seed photons were emitted by the accretion disc; both the components were multiplied by the {\tt phabs} component, which takes into account the photoelectric absorption by neutral matter in the interstellar medium.
We set the abundances to the values found by \cite{Wilms_00} and the photoelectric absorption cross-sections were set to the values obtained by \cite{Verner_96}.

Furthermore, in our analysis we assumed a distance to the source of 5.5 kpc \citep{Iaria_16} and a neutron star mass of 1.4 M$_{\odot}$.

\subsection{Renalysis of the {\it XMM-Newton/INTEGRAL} spectrum}
We re-analysed the {\it XMM-Newton/INTEGRAL} spectrum using the data in the energy ranges 0.6-2.0 keV for RGS12, 0.3-10 keV for MOS12, 2.4-10 keV for PN, 5-25 keV for JEM-X2, and 20-50 keV for ISGRI. 

We initially fitted the {\it XMM-Newton/INTEGRAL} data using the same self-consistent model used by \cite{Iaria_16}, in which the continuum emission was described as reported above. The inner disc temperature kT$_{\rm bb}$ and the disc-blackbody normalisation were left free to vary, as well as the photon index $\Gamma$, the electron temperature kT$_{\rm e}$ and the seed photons temperature kT$_{\rm seed}$.

To fit the emission line in the Fe-K region we adopted the self-consistent reflection model {\tt rfxconv} \citep{Kole_11} in which we kept fixed the iron abundance to the solar one and we left free to vary the ionisation parameter $\log \xi$ of the reflecting matter in the accretion disc and the reflection fraction rel\_refl. To take into account the relativistic smearing effects in the inner region of the accretion disc, we used the multiplicative {\tt rdblur} component, in which we kept fixed the outer radius at the value of 1000 gravitational radii ($R_g= G M/ c^2$), while left free to vary the inclination angle $\theta$ of the binary system, the inner radius R$_{\rm in}$ at which the reflection component originates and the power-law dependence of emissivity {\tt  betor}. The incident emission onto the accretion disc is provided by the Comptonisation component. We obtained best-fit results consistent with the results reported by \cite{Iaria_16} with a $\chi^2$(dof) of 2578(2247), but we observed large residuals between 3 and 4 keV. To fit these residuals, we added two relativistically broad lines with a fixed energy of 3.32 keV and of 3.9 keV (corresponding to the emission line of \ion{Ar}{xviii} and \ion{Ca}{xix} ions), respectively. We call this model
\begin{align*}
&{\tt Model \ 1}:\ \rm edge * edge * phabs *\{ diskbb + \\
			& rdblur * (gauss + gauss + rfxconv * nthComp[inp\_type=1])\}
\end{align*}
and show the best-fit results in the second column of \autoref{tab:res}. From this one we obtained a $\chi^2$(dof) of 2559(2245) and a $\Delta \chi^2=19$, with a F-test probability of chance improvement of 2.47$\times 10^{-4}$ (corresponding to a significance of $\sim$ 3.8$\sigma$) which suggests that the addition of both the two lines is statistically significant. The significance of the two lines are $\sim 4\sigma$ and $\sim 3\sigma$ for the \ion{Ar}{xviii} and \ion{Ca}{xix} emission line, respectively.

We observed a variation in all the parameters of the continuum, in particular we obtained a value of kT$_{\rm seed}$ larger than kT$_{\rm bb}$ suggesting that the seed photons are not emitted by accretion disc only but a possible contribution of photons emitted by the neutron star surface is present. For this reason, we fitted the data with 
\begin{align*}
{\tt Model \ 2}:\ \rm phabs*\{diskbb + nthComp[inp\_type=0]\},
\end{align*}
in which the {\tt inp\_type} parameter set to 0 indicates seed photons emitted by NS,  and we obtained a $\chi^2$(d.o.f.) of 2844(2252). 

Large residuals remain in the Fe-K region and between 3 and 4 keV; to fit those we added three {\tt diskline} \citep{Fabian_89} components for which we kept fixed the energies of two out of three lines at 3.32 and 3.9 keV, while the energy of the third one was left free to vary.
For the two lines with fixed energy, we tied the values of inclination angle, power law emissivity dependence, and black body radius to those of the third one. The outer radius of the reflecting region was set to 1000 R$\rm _g$ for each line. Finally, we imposed that the inner radius of the reflecting region had the same value of the inner radius of the accretion disc.
The best-fit results, obtained adding the three disk-lines, showed a $\chi^2$(d.o.f.) of 2640(2246), with a F-test probability of chance improvement of $1.13 \times 10^{-33}$. 
Large residuals were still present at 0.8 keV and 8.8 keV, for this reason we added two absorption edges fixing the energy threshold at 0.871 keV and 8.828 keV, associated with the presence of \ion{O}{viii} and \ion{Fe}{xxvi} ions, respectively. We called this model 
\begin{align*}
{\tt Model \ 3}:\ \rm & edge * edge * phabs * \{diskline + diskline + \\
					&diskline + diskbb + nthComp[inp\_type=0]\}.
\end{align*}

We obtained a $\chi^2$(d.o.f.) of 2523(2244), with an F-test probability of chance improvement of $1 \times 10^{-22}$ (corresponding to a significance much higher than 6$\sigma$) and a $\Delta \chi^2$ of 117 with respect to the previous model.
We obtained a significance of 19$\sigma$, 4$\sigma$ and 6$\sigma$ for the \ion{Ar}{xviii}, \ion{Ca}{xix} and \ion{Fe}{xxvi} emission lines, respectively.
The energy of the smeared emission line in the Fe-K region is $6.81 \pm 0.07$ keV, that is compatible within 3$\sigma$ with the rest frame value.
The best-fit results are shown in the third column of \autoref{tab:res}.

\begin{figure}
  \includegraphics[scale=0.6]{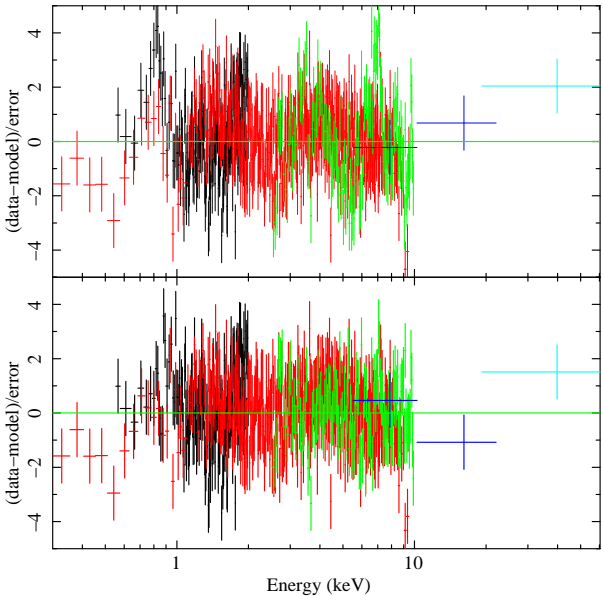}
  \caption{Comparison between residuals obtained adopting {\tt Model 2} (top panel) and {\tt Model 4} (lower panel). The black, red, green, blue and cyan colour indicate the RGS12, MOS12, PN, JEM-X2 and ISGRI data, respectively.}
       \label{fig:xmm_res}
  \end{figure}

Although {\tt Model 3} presented statistically good results, we decided to fit the data using also the self-consistent approach described above, which takes into account the reflection continuum. We kept fixed the outer radius in the {\tt rdblur} component at the value of 1000 R$\rm _g$ and the power law dependence of emissivity at the value of -2.5, consistently with the result of {\tt Model 3}.
In this case we imposed that the value of R$_{\rm in}$ is the same of R$_{\rm disc}$. 
Since the {\tt rfxconv} model does not account for the emission lines associated to ionised Argon and Calcium, we added two Gaussian components with energies fixed at 3.32 and 3.9 keV, respectively.
We call this model 
\begin{align*}
&{\tt Model \ 4}:\ \rm edge * edge * phabs * \{diskbb + \\
&rdblur * (gauss + gauss + rfxconv * nthComp[inp\_type=0])\}.
\end{align*}
Using this self-consistent model we obtained a $\chi^2$(d.o.f.) of 2550(2245), with a large improvement with respect to {\tt Model 2} and a significance of 6$\sigma$ for both \ion{Ar}{xviii} and \ion{Ca}{xix} emission lines.
We show the comparison between the residuals obtained using {\tt Model 2} and {\tt Model 4} in \autoref{fig:xmm_res}; the best-fit values of the parameters are shown in the fourth column of \autoref{tab:res}.
%
\begin{table}[!htbp]
\caption{Best-fit values of the spectral models for XMM-Newton data}
\label{tab:res}
\centering
\scriptsize
\begin{tabular}{lccccccccc}
\hline
& Model 1 & Model 3 & Model 4 & 
\\
Component \\
\hline
 
 {\tt Edge}\\
 E$_{\ion{O}{viii}}$ (keV)&
0.871 (fixed)&0.871 (fixed) &0.871 (fixed)&
\\

$\tau_{\ion{O}{viii}}$ &
$0.7 \pm 0.1$ & $0.7 \pm 0.1$ &$0.7 \pm 0.1$&
\\\\
 
{\tt Edge} \\
 E$_{\ion{Fe}{xxvi}}$ (keV)&
8.828 (fixed)&8.828 (fixed) &8.828 (fixed)&
\\
$\tau_{\ion{Fe}{xxvi}}$ &
$0.05 \pm 0.01$ &$0.04 \pm 0.01$ & $0.04 \pm 0.01$&
\\\\
 
{\tt phabs} \\
nH(10$^{22}$)&
 $ 2.44 \pm 0.04$&$2.40 \pm 0.02$ & $2.42 \pm 0.04$ & 
 & & & 
 \\ \\
 
{\tt diskbb} \\ 
kT$_{\rm in}$(keV) & 
 $0.43^{+0.02}_{-0.03}$ & $0.48^{+0.04}_{-0.12}$ & $0.46^{+0.03}_{-0.02}$ & 
\\

R$_{\rm disc}$ (km) & 
$21^{+4}_{-2}$& $20^{+8}_{-3} $ & $23^{+2}_{-3}$ &
\\

F$_{\rm bb}$\\
(10$^{-9} erg \ cm^{-2} \ s^-{1}$) &
& & 1.3 \\ \\

{\tt nthComp} \\
$\Gamma$ & 
$1.80^{+0.12}_{-0.07}$& $1.73^{+0.12}_{-0.07}$ & $1.79^{+0.03}_{-0.08}$ & 
\\
 
kT$_{\rm e}$(keV) & 
$2.9^{+0.3}_{-0.2}$&$2.6^{+0.3}_{-0.1}$ & $ 2.89^{+0.03}_{-0.02}$ & 
\\

kT$_{\rm seed}$(keV) &
$1.04^{+0.30}_{-0.20}$& $0.7^{+0.2}_{-0.4}$ & $0.7 \pm 0.1 $ & 
\\
 
norm & 
$0.20^{+0.04}_{-0.02}$& $ 0.08 \pm 0.02 $ & $0.09^{+0.02}_{-0.01}$ & 
\\

F$_{\rm Compt}$\\
(10$^{-9} erg \ cm^{-2} \ s^-{1}$) &
& & 2.4 \\ \\

{\tt diskline}\\
betor&  
-& $-2.5^{+0.2}_{-0.3} $ & - &
\\

$\rm \theta$ (deg)&
-& $35 \pm 3$&- &
\\

R$_{\rm in}$ (km) &
- & $20^{+8}_{-3} $ & - &
\\ \\

{\tt diskline}\\
E$_{\ion{Ar}{xviii}}$ (keV)&
-& 3.32 (fixed)& - &
\\

N$_{\ion{Ar}{xviii}}$  ($\times 10^{-4}$)&
-& $4.1 \pm 1.2$& -&
\\ \\

{\tt diskline}\\
E$_{\ion{Ca}{xix}}$ (keV)&
- & 3.90 (fixed)& - &
\\

N$_{\ion{Ca}{xix}}$  ($\times 10^{-4}$)&
-& $2.6^{+0.8}_{-1.1} $& -&
\\ \\

{\tt diskline}\\
 E$_{\ion{Fe}{xxv}}$ (keV)&
-& $6.81 \pm 0.07$& -&
\\

N$_{\ion{Fe}{xxv}}$  ($\times 10^{-4}$)&
-& $5.7 \pm 1.1 $& - &
\\ \\
 
{\tt rdblur} \\
Betor10 &
-2.8 (fixed) & - & -2.5 (fixed) &  
\\
 
R$_{\rm in}$ (km)&
$21^{+4}_{-2}$& - & $23^{+2}_{-3}$ & 
\\
 
$\theta$ (deg) & 
$36^{+2}_{-1}$& - & $38^{+7}_{-5}$& 
 \\ \\

{\tt gauss} \\
E$_{\ion{Ar}{xviii}}$ (keV)&
3.32 (fixed)& - & 3.32 (fixed)&
\\

N$_{\ion{Ar}{xviii}}$  ($\times 10^{-4}$)&
$3.9^{+1.1}_{-1.3}$ & - & $3.3^{+1.1}_{-0.9}$&
\\ \\

{\tt gauss} \\
E$_{\ion{Ca}{xix}}$ (keV)&
3.90 (fixed)& - & 3.90 (fixed)&
\\

N$_{\ion{Ca}{xix}}$  ($\times 10^{-4}$)&
$2.3^{+1.2}_{-1.4}$ & - & $1.5^{+1.0}_{-0.8}$ &
\\ \\
 
{\tt rfxconv} \\ 
rel\_refl & 
$0.09^{+0.03}_{-0.01}$ & - & $0.05^{+0.03}_{-0.01}$ & 
\\

$\log \xi$ & 
$2.72^{+0.2}_{-0.1}$& -  & $3.0^{+0.1}_{-0.3}$& 
\\

\hline

$\chi^2$(dof) & 
2559(2245) & 2523(2244) & 2550(2245) & 
\\

\hline
\end{tabular}
\tablefoot{\scriptsize Uncertainties are reported at 90\% confidence level. The spectral
 parameters are defined as in {\tt XSPEC}.
The units of the line normalisations are in $photons/cm^2 s^{-1}$.
$F_{\rm bb}$ and $F_{\rm Compt}$ and $F_{\rm bol}$ represent the unabsorbed fluxes in the (0.1-100) keV energy range associated with the accretion disc and the Comptonisation component, respectively.}
\end{table}
Finally, we find that the total unabsorbed flux is $3.8 \times 10^{-9}$ erg cm$^{-2}$ s$^{-1}$ in the 0.1-100 keV energy range; the corresponding luminosity is $1.5 \times 10^{37}$ erg s$^{-1}$.

\subsection{The {\it BeppoSAX} spectra}
We adopted the 0.12-4 keV, 1.8-10 keV, 7-25 keV and 15-100 keV
energy range for LECS, MECS, HPGSP and PDS spectra, respectively. 
Hereafter we call spectrum A, spectrum B and spectrum C, the spectra obtained from obs. A, B and C. 

Initially, we fitted the continuum emission with {\tt Model 2}. Assuming that N$_H$ does not change with respect to {\it XMM-Newton/INTEGRAL} observations and considering the low statistics of the \textit{BeppoSAX/LECS} data below 1 keV, we kept fixed the value of equivalent hydrogen column associated with the interstellar matter to $2.42 \times 10^{22}$ cm$^{-2}$, as obtained from best-fit values of {\tt Model 4} (see \autoref{tab:res}).
Fitting the data we obtained a $\chi^2$(d.o.f.) of 472(477), 650(502) and 532(481) for spectra A, B and C, respectively. The best-fit results are shown in the second, the third and the fourth column of \autoref{tab:sax}.
%
\begin{table*}
\caption{Best-fit values of the spectral models for BeppoSAX data}
\label{tab:sax}
\centering
\scriptsize
\begin{tabular}{lcccccc}
\hline
&\multicolumn{3}{c}{Model 2} & \multicolumn{3}{c}{Model 5} \\
& Obs A& Obs B& Obs C & Obs A& Obs B& Obs C\\
Component \\
\hline

{\tt phabs} \\
nH(10$^{22}$)&
 2.42 (frozen) & 2.42 (frozen) & 2.42 (frozen)& 
 2.42 (frozen) & 2.42 (frozen) & 2.42 (frozen) \\ \\
 
{\tt diskbb} \\ 
kT$_{\rm in}$(keV) & 
$0.54 \pm 0.02$ & $0.54 \pm 0.02$ & $0.48 \pm 0.06$ & 
$0.51 \pm 0.03$ & $0.49 \pm 0.02$ & $0.4 \pm 0.05$ \\

R$_{\rm disc}$ (km) & 
$9.8^{+0.08}_{-0.07} $ & $11.7 \pm 0.7$ & $8 \pm 2$ & 
$11^{+2}_{-1}$ & $12 \pm 1$ & $14 \pm 3$ \\

F$_{\rm bb}$\\
(10$^{-9} erg \ cm^{-2} \ s^-{1}$) &
& & &
0.5 & 0.4 & 0.2 \\ \\

{\tt nthComp} \\
$\Gamma$ & 
$2.44^{+0.04}_{-0.08}$& $2.24 \pm 0.03$ & $2.08 \pm 0.04$ & 
$2.3 \pm 0.1$ & $2.40^{+0.09}_{-0.08}$ & $2.09 \pm 0.04$ \\
 
kT$_{\rm e}$(keV) & 
$ >44$& $17 \pm 2$ & $34^{+13}_{-6}$& 
$>40$ & $28^{+20}_{-7}$ & $45^{+34}_{-13}$ \\

kT$_{\rm seed}$(keV) &
$1.02^{+0.03}_{-0.05}$ & $0.98 \pm 0.03$ & $0.73 \pm 0.06$ & 
$0.96 \pm 0.06$ & $1.48 \pm 0.07$ & $0.63 \pm 0.04$ \\
 
norm & 
$0.020^{+0.002}_{-0.001}$ & $0.026 \pm 0.002$ & $0.028 \pm 0.004$ & 
$0.018 \pm 0.002$ & $0.079 \pm 0.003$ & $0.035^{+0.002}_{-0.004}$\\
 
F$_{\rm Compt}$\\
(10$^{-9} erg \ cm^{-2} \ s^-{1}$) &
& & &
0.95 & 1.4 & 1.0 \\ \\

{\tt rdblur} \\
Betor10 &
- & - & - & 
 -2.5 (frozen) & -2.5 (frozen) & -2.5 (frozen) \\
 
R$_{\rm in}$ (km) &
- & - & - & 
24 (frozen) & $<39$ & $<50$ \\
 
$\theta$ (deg) &
 - & - & - &
38 (frozen) & 38 (frozen) & 38 (frozen) \\ \\
 
{\tt rfxconv} \\ 
rel\_refl & 
- & - & - & 
$0.09 \pm 0.04$ & $0.4 \pm 0.1$ & $0.2 \pm 0.1$\\

$\log \xi$ & 
- & - & - & 
3.14 (frozen) & $2.32 \pm 0.04$ & $2.38^{+0.33}_{-0.05}$ \\

F$_{\rm bol}$\\
(10$^{-9} erg \ cm^{-2} \ s^-{1}$) & 
& & &
1.5 & 2.1 & 1.3 \\

\hline

$\chi^2$(dof) & 
472(477)& 650(502)& 532(481)&  
449(476) & 558(499) & 482(478) \\

\hline
\end{tabular}
\tablefoot{\scriptsize Uncertainties are reported at 90\% confidence level. The spectral
 parameters are defined as in {\tt XSPEC}.\\
 $F_{\rm bb}$, $F_{\rm Compt}$ and $F_{\rm bol}$ represent the unabsorbed fluxes in the (0.1-100) keV energy range associated with the accretion disc, the Comptonisation component and the total emission, respectively.}
\end{table*}

\begin{figure*}
  \centering
  \includegraphics[scale=0.4]{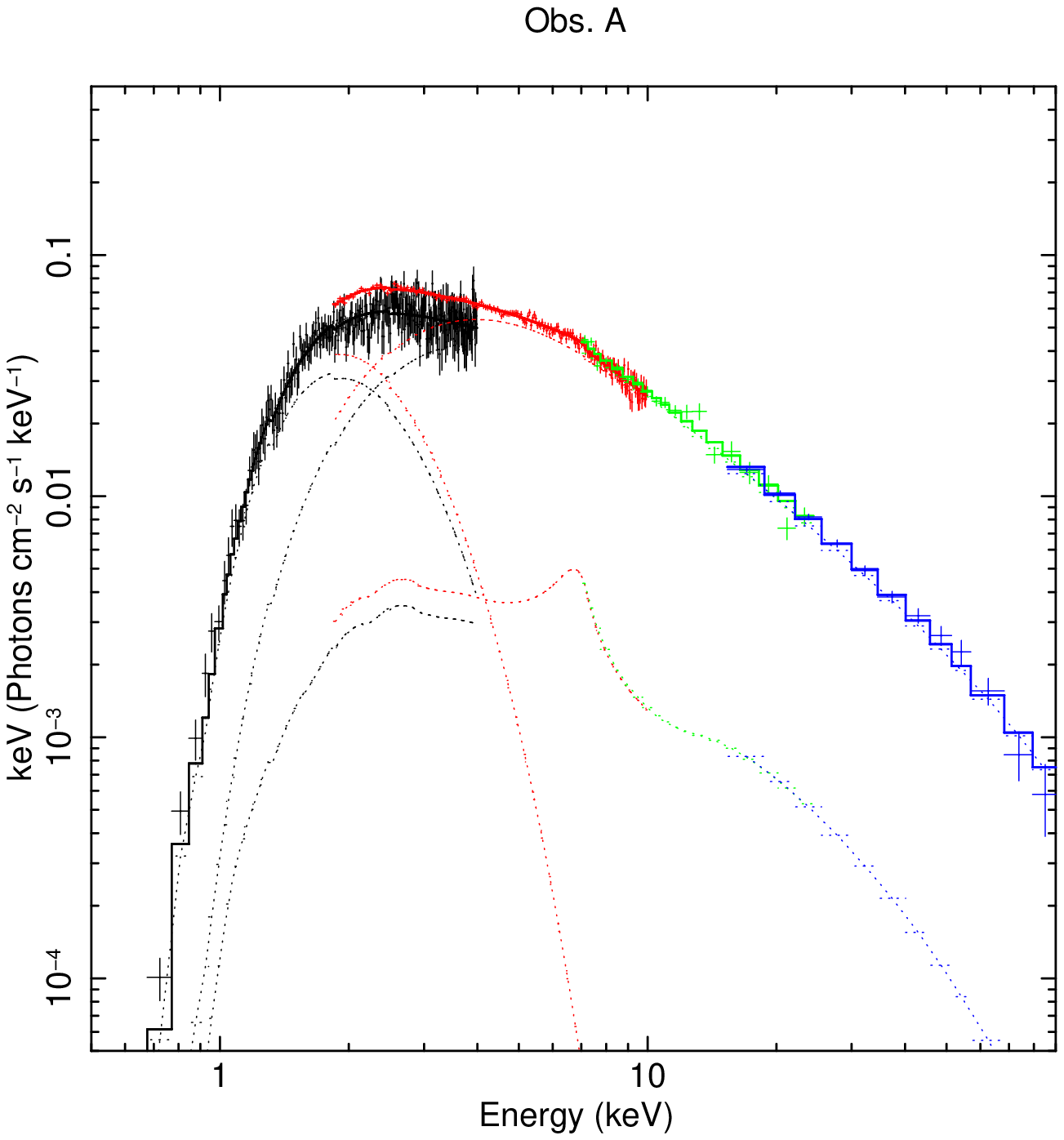}
  \includegraphics[scale=0.4]{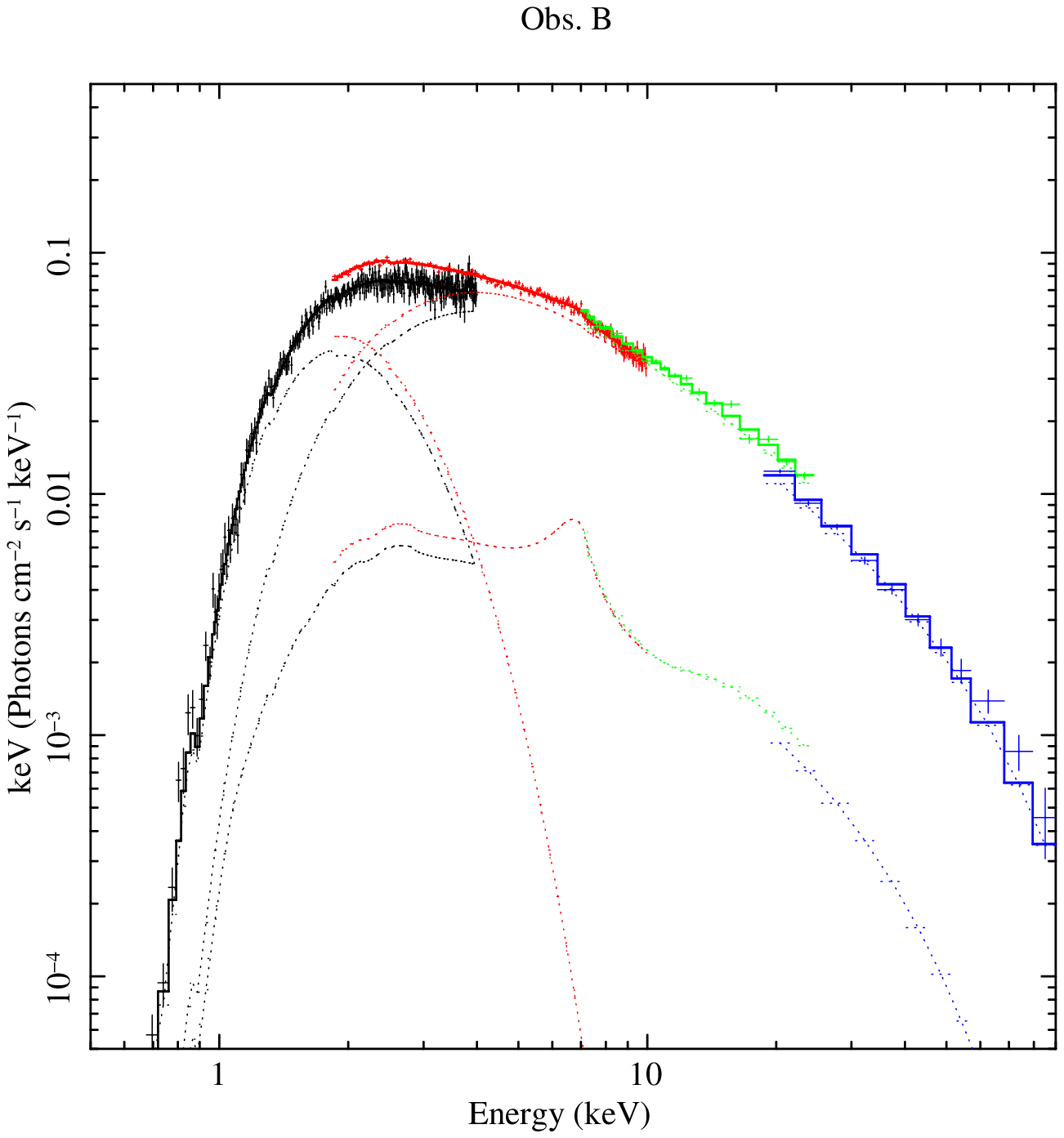}
  \includegraphics[scale=0.4]{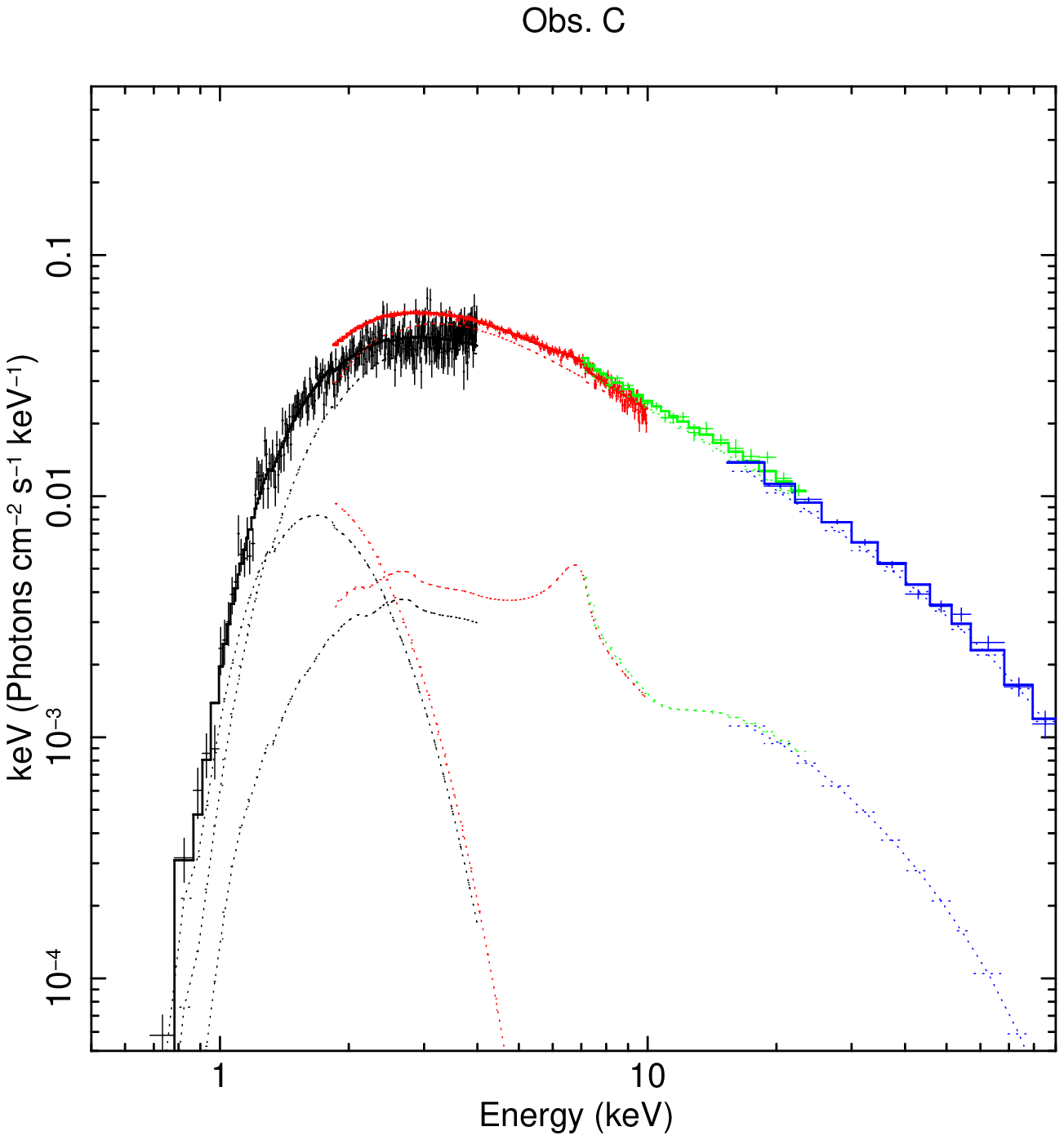}
  \caption{The unfolded spectra of the three \textit{BeppoSAX} observation fitted adopting {\tt Model 5}. The black, red, green and blue data represent the LECS, MECS23, HPGSP and PDS spectra, respectively.}
       \label{fig:pha2}
  \end{figure*}

\begin{figure*}
\centering
  \includegraphics[scale=0.4]{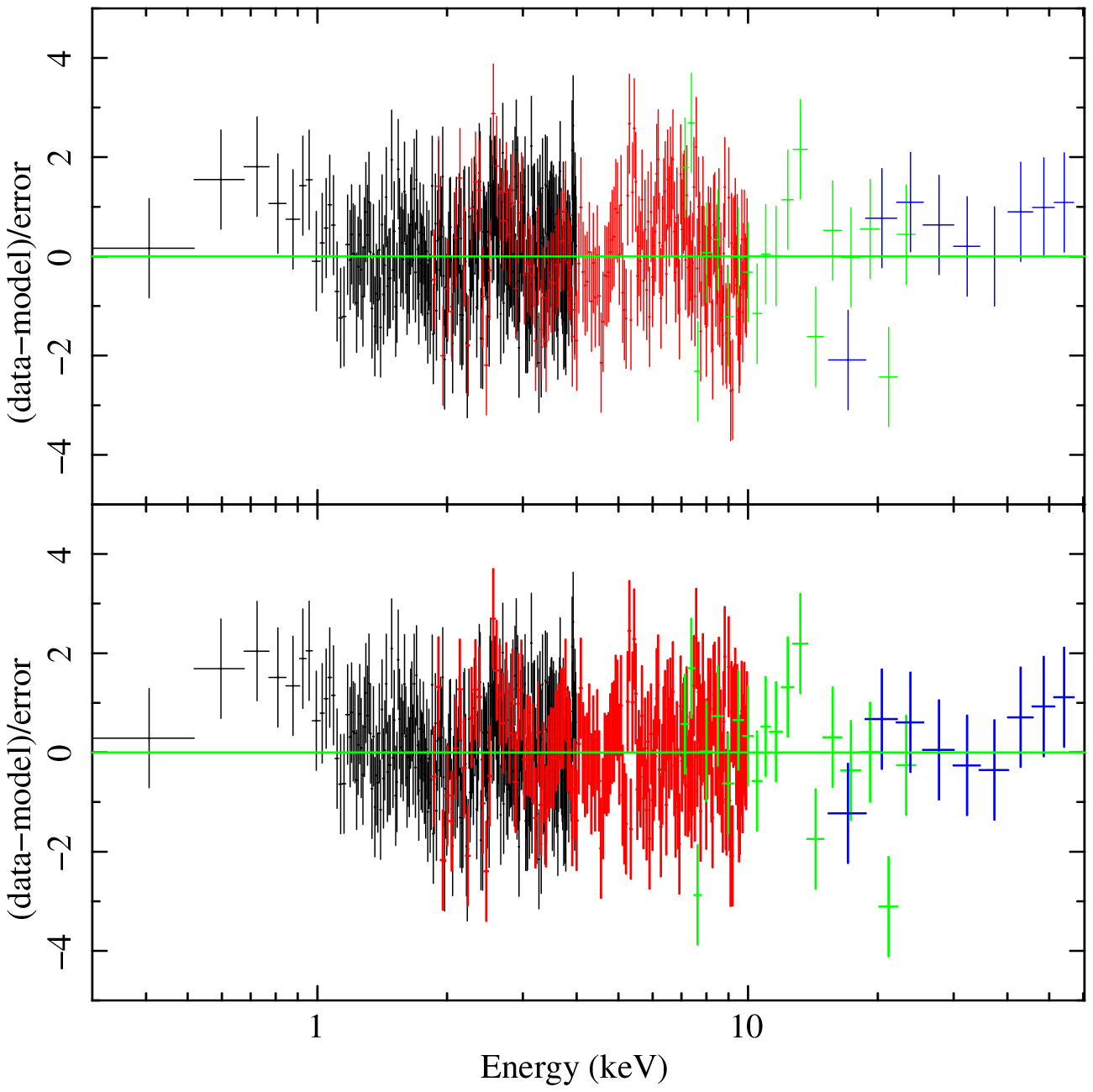}
  \includegraphics[scale=0.4]{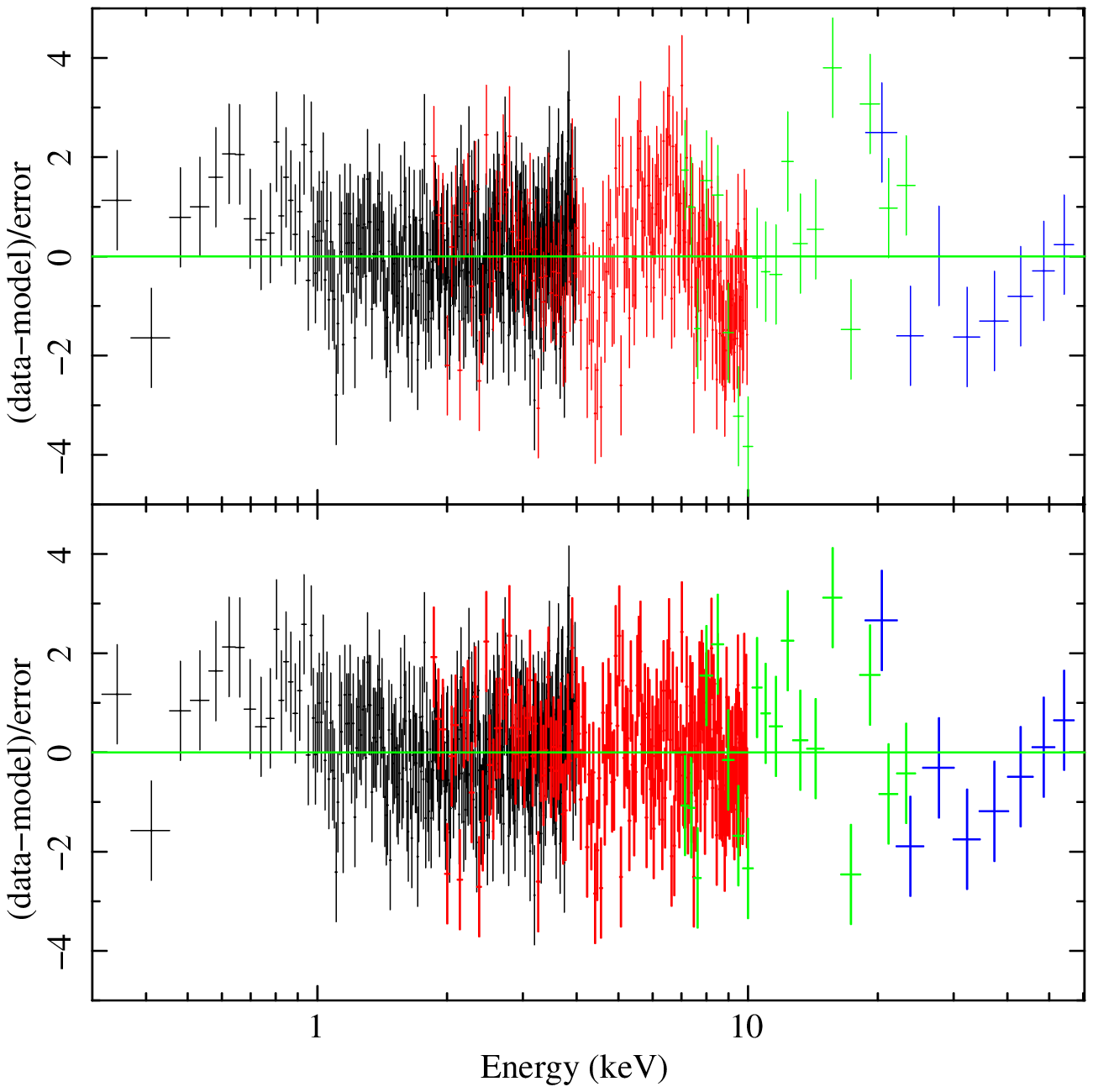}
  \includegraphics[scale=0.4]{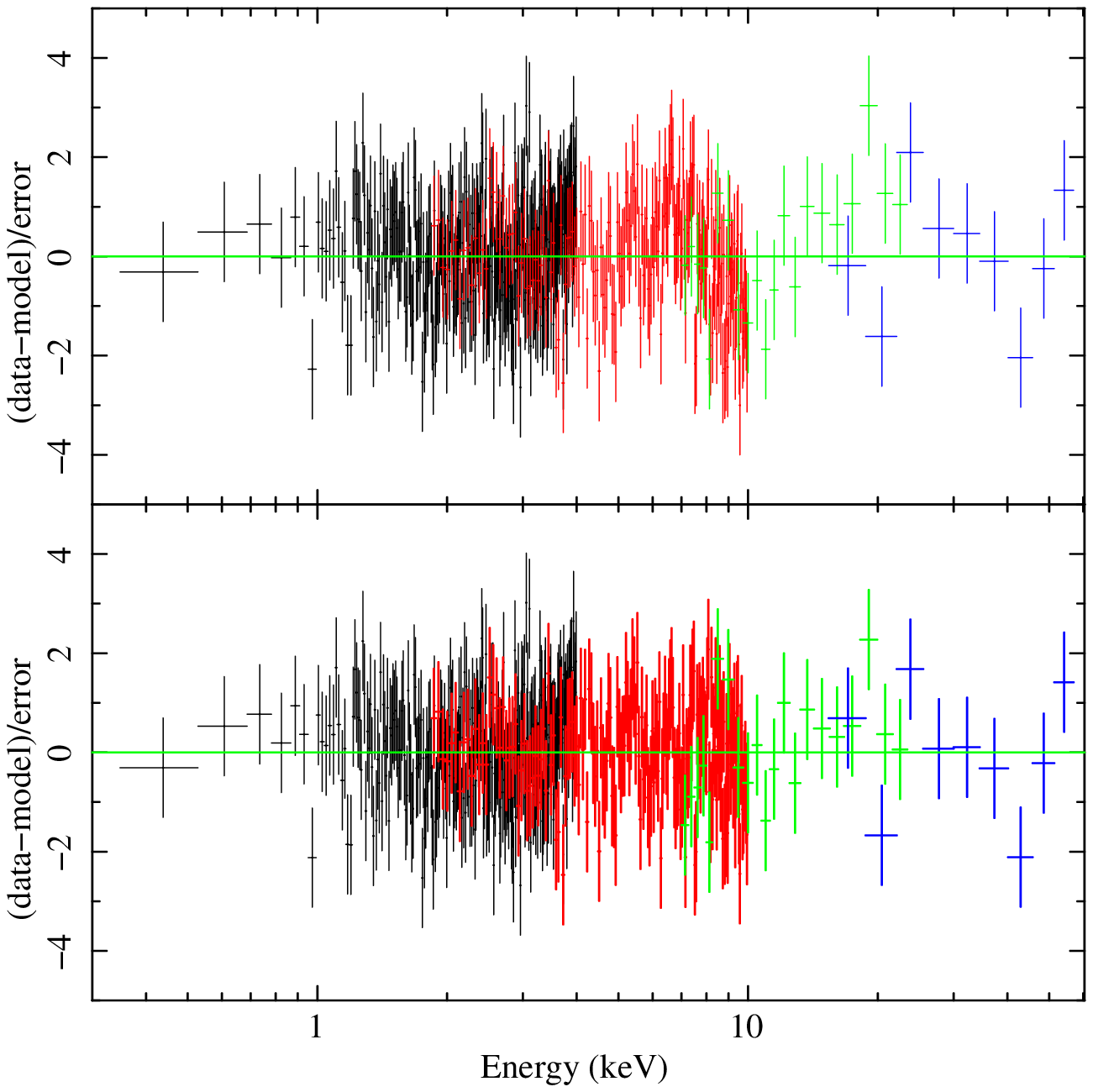}
  \caption{Comparison between residuals obtained adopting {\tt Model 2} (top panels) and {\tt Model 5} (bottom panels) for Obs. A (on the left), Obs. B (in the middle) and Obs. C (on the right). The black, red, green and blue points represent the LECS, MECS23, HPGSP and PDS data, respectively.}
    \label{fig:res}
  \end{figure*}

\begin{figure}
\centering
  \includegraphics[scale=0.6]{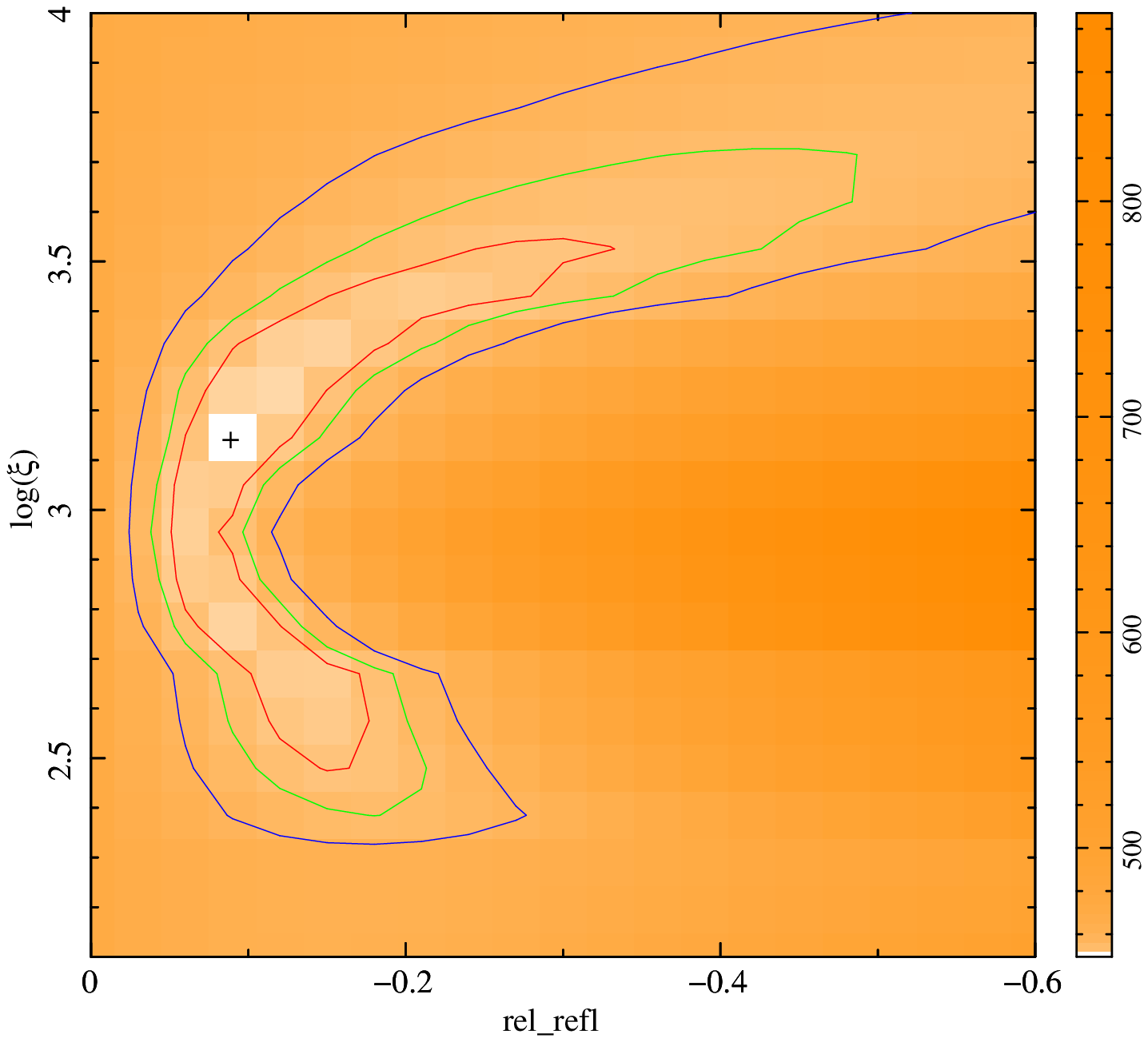}
  \caption{Contour plot of $\chi^2$ changes for simultaneous variation of $\log  \xi$ and rel\_refl parameters for spectrum A. The red, green and blue lines indicate the contours at 68\%, 90\% and 99\% confidence level, respectively. The  black cross indicates the best-fit values of $\log \xi=3.14$ and $rel\_refl=0.1$. The orange-scale indicates as the value of $\chi^2$ changes in the grid.}
    \label{fig:contA}
  \end{figure}
We observed that some residuals are present in the Fe-K region of the spectra A and C and these are slightly larger for spectrum B. For this reason we added a smeared reflection component to {\tt Model 2}. We adopted the Compton reflection model described for the analysis of the \textit{XMM-Newton/INTEGRAL} spectrum. The inclination angle was kept fixed to $\ang{38}$ as obtained from {\tt Model 4} (see \autoref{tab:res}). In this case, the inner radius R$_{\rm in}$, the ionisation parameter $\log \xi$ and the reflection fraction rel\_refl were left free to vary.
We call this model
\begin{align*}
 &{\tt Model \ 5}:\\
 &  \rm phabs * \{diskbb + rdblur * rfxconv * nthComp[inp\_type=0]\}.
\end{align*}

For obs. B we obtained a $\chi^2$(d.o.f.) of 558(499), the F-test probability of chance improvement for the addition of the reflection component is $2.0 \times 10^{-16}$. The Fe-K region of the spectrum does not show significant residuals anymore.
The best-fit values of the parameters are shown in the sixth column of \autoref{tab:sax}. The unfolded spectrum and the residuals are shown in the central panels of \autoref{fig:pha2} and \autoref{fig:res}, respectively.

Subsequently, we adopted the {\tt Model 5} to fit spectra A and C.
For spectrum C, we obtained a $\chi^2$(d.o.f.) of 482(478) and a F-test probability of chance improvement of $3.1 \times 10^{-10}$ (significance higher than 6$\sigma$). The ionisation parameter $\log \xi$ and the relative reflection normalisation rel\_refl are $2.38^{+0.33}_{-0.05}$ and $0.2 \pm 0.1$, respectively, and they are compatible within 90\% c.l. with the values obtained for spectrum B.
The best-fit values of the parameters for spectrum C are shown in the seventh column of \autoref{tab:sax}.

For spectrum A, we needed to fix the value of the ionisation parameter in order to led the fit to convergence. We chose the best value $\xi=$3.14 obtained from the contour plot, shown in \autoref{fig:contA}. This two-dimensional distribution of the $\chi^2$ as a function of the ionisation and the reflection amplitude was obtained varying simultaneously $\log (\xi)$ between 2.1 and 4 and rel\_refl between -0.6 and -0.01 through the {\tt steppar} tool in {\tt XSPEC}. Moreover, we kept fixed the inner radius of the reflection region to the value of 24 km, because the fit did not reach any constrains for this one and the $\chi^2$ was insensible to the variation of this parameter. The value of 24 km was obtained from {\tt Model 5} leaving R$_{\rm in}$ free to vary and keeping fixed all the other parameters.
We obtained a $\chi^2$(d.o.f.) of 449(476) and the F-Test gives a probability of chance improvement of $1.1 \times 10^{-6}$, corresponding to a significance of $\sim 4.9\sigma$.

The best-fit values of the parameters for spectrum A are shown in the fifth column of \autoref{tab:sax}.
The unfolded spectrum and the residuals are shown in the left and right panels of \autoref{fig:pha2} and \autoref{fig:res} for spectra A and C, respectively.

We found that the total unabsorbed flux $1.5 \times 10^{-9}$ erg cm$^{-2}$ s$^{-1}$ for spectrum A, $2.1 \ \times 10^{-9}$ erg cm$^{-2}$ s$^{-1}$ for spectrum B, and $1.3 \times 10^{-9}$ erg cm$^{-2}$ s$^{-1}$ for spectrum C.
Finally, the corresponding luminosities are $5.5 \times 10^{36}$ erg s$^{-1}$, $7.5\times 10^{36}$ erg s$^{-1}$ and $4.8 \times 10^{36}$ erg s$^{-1}$ for spectra A, B and C, respectively.

\section{Discussion}
We re-analysed the observations of the NS-LMXB 4U 1702-429 taken with
{\it XMM-Newton/INTEGRAL} in March 2010 and we analysed three {\it BeppoSAX} observations performed in February 1999, August 2000 and September 2000. The broadband analysis indicates that the addition of a smeared reflection component to the spectral model is statistically significant in the case of {\it XMM-Newton/INTEGRAL} observation and observations B and C, and it is marginally significant for observation A.

The smeared reflection component allowed to constrain the inclination angle $\theta$ of the source, that is $38^{+7}_{-5}$ degrees using the {\it XMM-Newton/INTEGRAL} spectrum. 

The 0.1-100 keV unabsorbed luminosity is a factor by two/three larger during the {\it XMM-Newton/INTEGRAL} observation than that obtained from the three {\it BeppoSAX} observations (see the section above). Moreover, the electron temperature is $2.89^{+0.03}_{-0.02}$ keV for the {\it XMM-Newton/INTEGRAL} spectrum  while it is $40^{+93}_{-14}$ keV, $28^{+20}_{-7}$ keV and $45^{+34}_{-13}$ keV for spectra A, B and C, respectively. This suggests that the source was in a SS during the {\it XMM-Newton/INTEGRAL} observation and in a HS during the three {\it BeppoSAX} observations.

From {\tt Model 4} we obtain R$_{\rm disc} = 23^{+2}_{-3}$ km for {\it XMM-Newton/INTEGRAL} spectrum, while R$_{\rm disc}$ is $11^{+2}_{-1}$ km, $12 \pm 1$ km and $14 \pm 3$ km for spectra A, B and C, respectively, adopting {\tt Model 5}. We applied the correction factor to convert the inner radius R$_{\rm disc}$ values into the realistic inner radius $r_{\rm disc}$ values. The relation between these two radii is $r_{\rm {disc}} \sim f^2 R_{\rm disc}$, where the colour correction factor $f \simeq 1.7$ for a luminosity close to 10\% of Eddington luminosity \citep{Shimu_95}. We find that the $r_{\rm {disc}}$ values are $66^{+5}_{-9}$ km, $32^{+6}_{-3}$ km, $35 \pm 3$ km and $40 \pm 9$ km for {\it XMM-Newton/INTEGRAL} observation and for spectra A, B and C, respectively.
The $r_{\rm disc}$ values are compatible with each other within 90\% c.l. for the three \textit{BeppoSAX} spectra, while the value obtained from {\it XMM-Newton/INTEGRAL} spectrum is a factor about two larger. We note that the behaviour of the disc blackbody radius is similar to what obtained by \cite{DiSalvo_15} for the prototype atoll source 4U 1705-44. In that case the authors observed an inner radius of the blackbody component close to 11 km in the HS and about 33 km in the SS. Then, also the obtained values of $r_{\rm {disc}}$ seem to suggest that the spectrum of 4U 1702-429 is in a SS during the {\it XMM-Newton/INTEGRAL} observation, while it is in a HS during the three {\it BeppoSAX} observations.

This result could be explained assuming that the inner region of the accretion disc is occulted by an optically thick corona during the {\it XMM-Newton/INTEGRAL} observation. 
To verify the proposed scenario we estimated the optical depth $\tau$ of the Comptonising cloud, using the relation provided by \cite{Zdi_96}:
\begin{equation*}
\Gamma= \left[\dfrac{9}{4} + \dfrac{1}{\tau \left(1+ \frac{\tau}{3} \right) \frac{kT_e}{m_ec^2}} \right]^ {1/2} - \dfrac{1}{2}.
\end{equation*} 
We found that $\tau = 11.9^{+0.2}_{-0.8}$ during the {\it XMM-Newton/INTEGRAL} observation while $\tau=1.5^{+0.4}_{-0.9}$, $\tau = 1.8^{+0.3}_{-0.7}$ and $\tau = 1.7^{+0.5}_{-0.6}$ for spectra A, B and C, respectively.
Our results suggest that the corona is optically thick during the {\it XMM-Newton/INTEGRAL} observation, while the optical depth is much lower during the {\it BeppoSAX} observations. The obtained values of $\tau$ support the idea that the innermost region of the system is embedded in an optically thick corona in the SS, while the optically thin corona observed in the spectra A, B and C allows to look through at the innermost region.

Since the inner radius is between 30 km and 50 km (for spectra A, B and C) we might suppose that the accretion disc is truncated and it does not reach the NS surface. This hypothesis is endorsed by the general idea that in a HS the accretion disc is truncated and, in the region near the NS, the hot inner corona can be identified with the boundary layer, whose interactions with the colder accretion disc cause the emptying of the innermost region \citep{Roza_00}. Then the flow and the boundary layer become optically thinner and the thermal emission can be dominated by the neutron star, as suggested by the smallest value of the inner radius of the accretion disc obtained in HS \citep{Barret_02}.

The two observed spectral states are identified by changes in the electron corona. We note that for {\it XMM-Newton/INTEGRAL} spectrum the electron temperature kT$_e$ is about tens of times smaller.
Since $\tau \propto N_e R$, where $R$ is the size of the Comptonising region, an increase in the optical depth ($\tau$ is ten times larger during the {\it XMM-Newton/INTEGRAL} observation) can be associated with an increase in the electron density $N_e$ of the corona. Then, an increase in the electron density entails an increase in the number of the electrons in the cloud, which is consistent with the observed increase in the flux associated with Comptonised component.

Assuming the matter almost totally ionised in the accretion disc, we also estimated the electron density $n_e$ of the reflecting skin using the relation $\xi = L_x/(n_er^2)$, where $L_x$ is the unabsorbed incident luminosity between 0.1 and 100 keV, $\xi$ is the ionisation parameter, and $r$ is the inner radius of the disc where the reflection component originates. We find that for the {\it XMM-Newton/INTEGRAL} spectrum $n_e= (1.6^{+1.1}_{-0.5}) \times 10^{21}$ cm$^{-3}$; on the other hand, since we found only an upper limit for the reflecting radius for \textit{BeppoSAX} observations, we obtain only a lower limit for the electron density: it is $1.7 \times 10^{21}$ cm$^{-3}$ for spectrum A, $1.6 \times 10^{21}$ cm$^{-3}$ for spectrum B and $0.6 \times 10^{21}$ cm$^{-3}$ for spectrum C.
Therefore, we estimated the seed-photon radius R$_{\rm seed}$ using the relation $R_{\rm seed}= 3 \times 10^4 d [F_{\rm Compt} /(1+y)]^{1/2} (kT_{\rm seed})^{-2}$ \citep{Zand_99}, where $d$ is the distance to the source in units of kpc, $y= 4kT_e\ max[\tau,\tau^2]/(m_ec^2)$ is the Compton parameter, kT$_{\rm seed}$ is the seed-photon temperature in units of keV and, finally, $F_{\rm Compt}$ is the bolometric flux of the Comptonisation component. 
Using the best-fit values (reported in \autoref{tab:res} and \autoref{tab:sax}), we find that $R_{\rm seed}$ is $8 \pm 2$ km from the {\it XMM-Newton/INTEGRAL} spectrum, and it is $4.3^{+0.7}_{-0.6} $ km, $2.2^{-0.4}_{-0.3}$ km and $10 \pm 2$ km from the \textit{BeppoSAX} spectra A, B and C, respectively. Although we can not exclude that seed photons originate in the inner accretion disc, this suggests that during observations C and {\it XMM/INTEGRAL} the seed photons were mainly emitted from the NS surface, while the smallest values obtained from spectra A e B can be explained assuming that only an equatorial region of the NS surface injects photons in the electron corona.

Typical values of the reflection amplitude $\Omega/2\pi$ for NS-LMXB atoll sources are within 0.2-0.3 \citep{MatrEXO_17,DiSalvo_15,Egron_13,Dai_10}. However, smaller reflection fractions are also observed for other systems, such as Ser X-1 \citep{Matranga_17,Cackett_10} and 4U 1820-30 \citep{Cackett_10} and for AMSP SAX J1748.9-2021 \citep{Pintore_16}. We found typical values of $\Omega/2\pi$ for \textit{BeppoSAX} observation B and C ($0.4 \pm 0.1$ and $0.2 \pm 0.1$, respectively), whilst we obtained a reflection fraction of $0.05^{+0.03}_{-0.01}$ for \textit{XMM-Newton/INTEGRAL} spectrum, that agrees with the value obtained by \cite{Iaria_16}. Since this parameter is a measure of the solid angle subtended by the reflector as seen from the Comptonising corona, our results suggest a slightly different geometry of the electron cloud between SS and HS. In particular, the value obtained for \textit{XMM-Newton/INTEGRAL} observation could indicate a small subtended angle due, for instance, to a less efficient interaction between the corona and the accretion disc or a different geometry of the cloud with respect to a central spherical corona contiguous to an outer accretion disc (which is a typical representation). Finally, we found that rel\_refl$=0.09 \pm 0.04$ for spectrum A, which is compatible with the value obtained for spectrum C, although the reflection component is unconstrained in this case and this does not represent a conclusive result.

In the case of this source the variation in the spectral state is likely caused by the changes occurred in the Comptonising corona, as suggested by \cite{Barret_02} for 4U 1705-44. Indeed, the spectral component whose parameters show significant variation between {\it XMM-Newton/INTEGRAL} analysis and {\it BeppoSAX} analysis is the Comptonisation (see \autoref{tab:res} and \autoref{tab:sax}).

In conclusion, we do not observe in the {\it BeppoSAX} spectra emission lines and absorption edges as detected in the {\it XMM-Newton/INTEGRAL} spectrum because of the smaller effective area and worse energy resolution of \textit{BeppoSAX/MECS} with respect to the \textit{XMM/EPIC-PN}.

\subsection{BH-systems comparison}
Binary systems hosting a neutron star show similar spectral characteristics with respect to black hole systems, and it would be very hard to distinguish between these two kind of systems looking only at their spectra. In the BH-candidates systems the accreting matter may free fall into the compact object because close to the event horizon the gravitational force predominates the pressure forces. On the other hand, in the NS systems, the free fall of accreting matter is slowed down by the presence of the solid surface emitting a large part of the total flux and therefore radiation pressure forces should be predominant. Generally this means a mass accretion rate higher than for a NS system and possibly a hotter Comptonized component \citep{DiSalvo_06}. 

In BH systems a correlation seems to be present between the softness (or hardness) of the spectrum and the value of the photon index, in particular the steeper the power law is, the softer is the spectrum \citep{Steiner_16}. For NS systems this correlation seems to be not always present: for example in the case of 4U 1702-429, for the {\it BeppoSAX} data the photon index is higher than for the {\it XMM-Newton/INTEGRAL} spectrum, although the BeppoSAX spectra are harder than this one. Furthermore, as for BH systems the spectral variation is led by the variation in the flux of the seed photons in the corona and then in a variation in the flux of the outgoing Comptonised photons \citep{DelSanto_08}.

The emission lines observed in the spectra of NS systems result broadened and skewed as lines detected in BH systems spectra, suggesting that in both the cases they are produced in the innermost region of the accretion disc, where the effects of the gravitational field of the compact object is stronger. However, both in HS and SS of NS systems, the broadening of the line is not as extreme as in the case of some black hole X-ray binaries or AGNs \citep[see e.g.][]{Reis_09,Fabian_09}.

The reflection component has a similar behaviour in BH and NS  
systems and it results more evident in the SS than in the HS, at least in the standard X-ray range (2-10 keV), with the presence of stronger features (see e.g \cite{DiSalvo_15,Egron_13,DiSalvo_09} for NS systems and \citet{Steiner_16,Zdi_02} for BH systems).
In the case of the source 4U 1702-429 the comparison of the reflection features between SS and HS is difficult because of the low statistics at higher energy for {\it BeppoSAX} data, which do not allow to observe the same emission lines observed in {\it XMM-Newton/INTEGRAL} spectrum. We obtained incompatible values of the reflection fraction and the ionisation parameter, which suggest that the geometry of the system and the physical characteristics of the electron corona had a variation between the two observed spectral states.

\section{Conclusions}
We have performed a detailed broadband spectral analysis of the source 4U 1702-429 in the 0.3-60 keV energy range using an {\it XMM-Newton/INTEGRAL} observation and using the data of three observations collected by \textit{BeppoSAX} in the 0.1-100 keV energy range. Using a self-consistent reflection model, the main results of our analysis are the following:
\begin{itemize}
\item the reflection fraction and the ionisation parameter of the reflection component are incompatible between the {\it XMM-Newton/INTEGRAL} spectrum and the {\it BeppoSAX} observations. Moreover we observe an optically thin electron cloud with an electron temperature larger than 30 keV in the \textit{BeppoSAX} spectra and a colder ($\sim$3 keV) optically thick corona for the \textit{XMM-Newton/INTEGRAL} one. Furthermore, we find that the total unabsorbed bolometric flux is only twice larger in the case of the \textit{XMM-Newton/INTEGRAL} observation with respect to \textit{BeppoSAX} observations;
\item in the \textit{XMM-Newton/INTEGRAL} spectrum we detect the presence of three emission lines, due to the fluorescence emission from \ion{Ar}{xviii}, \ion{Ca}{xix} and \ion{Fe}{xxv}, and two absorption edges identified as the presence of \ion{O}{viii} and \ion{Fe}{xxvi} in the accretion disc;
\item for the \textit{BeppoSAX} observations the best-fit parameters describe a physical scenario of a source in a HS, whilst the \textit{XMM-Newton/INTEGRAL} spectrum could be associated with a SS;
\item 
the inner radius of the accretion disc seems to be smaller in the HS (i.e. the disc is closer to the NS surface). This is probably due to the presence of an optically thinner corona than in SS, which allows to observe the emission from the innermost region of the system. In particular, we might observe a higher contribution from the boundary layer near the NS surface, which is shielded instead from the optically thick corona during SS. This suggests that in SS the thermal component is probably well-fitted by a disc multicolour blackbody, while in HS two blackbody component might be a better description of the soft component: the first one to describe the accretion disc emission and the second one for the boundary layer emission \citep[see e.g.][]{Armas_17}.
\end{itemize}
Further broadband observations are needed to confirm our scenario. 
  
\begin{acknowledgements}
Part of this work is based on archival data and software provided by
the Space Science Data Center - ASI.
This research has made use of data and/or software provided by the High Energy Astrophysics Science Archive Research Center (HEASARC), which is a service of the Astrophysics Science Division at NASA/GSFC and the High Energy Astrophysics Division of the Smithsonian Astrophysical Observatory.
The authors acknowledge financial contribution from the agreement ASI-INAF n.2017-14-H.0. and from the agreement ASI- INAF I/037/12/0 and the support from the HERMES Project, financed by the Italian Space Agency (ASI) Agreement n. 2016/13 U.O.
Part of this work has been funded using resources from the research grant “iPeska” (P.I. Andrea Possenti) funded under the INAF national call Prin-SKA/CTA approved with the Presidential Decree 70/2016.
S.M.M. thanks the research project ``Stelle di neutroni come laboratorio di Fisica della Materia Ultradensa: uno studio multifrequenza" financed by Regione Autonoma della Sardegna (scientific project manager prof. Luciano Burderi) in which part of this work was developed. 
The authors thank Dr. Milvia Capalbi for her kindly scientific and technical support.
The authors would like to thank the anonymous referee for his/her helpful comments.
\end{acknowledgements}

\bibliographystyle{aa}
\bibliography{biblio}

\end{document}